\documentclass[preprint,aps]{revtex4}
\usepackage{graphicx}
\usepackage{rotating}
\usepackage{bm}
\usepackage{subfigure}

\begin{document}
\title{Low temperature heat capacity of amorphous systems: physics at nano-scales }
\author{Pragya Shukla}
\affiliation{Department of Physics, Indian Institute of Technology, Kharagpur-721302, India  }
\date{\today}

\widetext

\begin{abstract}

Contrary to previous studies of boson peak, we analyze the density of states and specific heat contribution of dispersion forces in an amorphous solid of nano-scales ($\sim 3 nm$). Our analysis indicates a universal semi-circle form of the average density of states in the bulk of the spectrum along with a super-exponentially increasing behavior in its edge. The latter in turn leads  to a specific heat, behaving linearly below $T < 1^o \; {\bf K}$ even at nano-scales, and,  surprisingly agreeing with the experiments although the latter are carried out at macroscopic scales. The omnipresence of dispersion forces at microscopic scales indicates the application of our results to other disordered materials too.

\end{abstract}

\maketitle
.


\section{Introduction}

At low temperatures, structurally and orientationally disordered solids such as dielectric and metallic glasses, amorphous polymers and even crystals are experimentally observed  to exhibit many universalities \cite{zp, stephen, plt}. Initially the origin of this behavior was attributed to the tunnelling two level  systems (TTLS)  intrinsic to disordered state \cite{and,phil}. The existence of TTLS entities however in a wide range of materials is not  well-established (besides many other issues e.g. see \cite{galp, ram, rpjr,lg1, lg2,yl, pcrrr, liu, zink, que, agl, ang}). Previous attempts in search of alternative explanations led to introduction of many new models e.g. soft potential model \cite{spm,buch,paras,gure},  heterogeneous elasticity theory \cite{schi1, schi2}, Euclidean random matrices \cite{grig},  effective medium theories and jamming approach (e.g. \cite{emt, degi} and references therein), correlations at microscopic scales  \cite{vdos,du, ell3,mg} and several others \cite{yu,vlw,misha,burin,Seth,vl,zac} but almost none of them analyzed the existence of universalities at nanoscales \cite{ram}. With increasing industrial significance of nano materials, it becomes imperative to seek the information e.g whether these glass peculiarities emerge only at macrosizes (with experimental sample size typically $\sim 1 \;cm^3$) or  exist at nanoscales too? and if not then when and how much they deviate? The primary focus of the present work is to bridge this information gap by an analysis of the density of states and low temperature specific heat  at these scales.

The appearance of similar universal features in a wide range of disordered  systems strongly suggest the behavior  to originate in fundamental type of interactions. This intuitively motivates us to consider the inter-molecular interactions, more specifically the Vanderwaal (VW) forces among the molecules of the amorphous solid  as the root cause for the behavior. The intuition comes from the omnipresence of these forces  in all condensed phases thus making them promising candidates to decipher the experimentally observed universality.

Amorphous solids possess a rich and varied array of short to medium range topological order which originates from their chemical bonding and molecular interactions. The bonding interactions at short range (upto 1st and 2nd nearest neighbor molecular  distances), are dictated by the valance force fields (e.g covalent or ionic, also acting as mechanical constraints)  but the medium range order (typically of the order of $10-30\; \AA$),  is governed by the VW interactions \cite{phil2, ell1, sme, b2008, b1999, b1994}. Intuitively these varying structures at different length scales are expected to result in materials with very different physical properties.  Any theory of experimentally observed universality therefore must take into account the microscopic structure of the solid and explain the emergence of system-independence of the physical properties at large scales (as the related experiments are usually carried out on macroscopic sizes). The point to ponder is whether the observed universalities are an emergent phenomenon due to interactions at larger scale or even the molecular interactions within the short length scales (of medium range order)  can give rise to them on their own. The query arises because the experiments on nano-materials exhibit  physical and chemical properties that are significantly different from the corresponding properties of the bulk materials (e.g. see \cite{nano1} and references therein).  It is therefore relevant to seek, identify and analyze the behavior of the  subunits (nanoscale structures, later referred as basic blocks)  of the macroscopic glass solid.

Our primary focus in the present work is to understand the low temperature behavior of the specific heat for  one such subunit i.e a nanosize glass sample. Based on the standard partition function approach, this requires a prior knowledge of the density of states and therefore its Hamiltonian ${\mathcal H}$. The latter, although acting in an infinite dimensional Hilbert space, can be represented by a large but finite-size matrix in a physically motivated basis, with the size dictated by energy-range of interactions. Due to complexity of the many-body interactions within the block, however, the matrix elements can not be exactly determined and can at  best be described by their distribution over an ensemble of the replicas of $H$-matrix. Intuitively, the form of the distribution or the distribution parameters are expected to be system specific.  The experimentally observed universality at low temperatures however suggests that the theoretical  formulation should not depend on any system-specific details.  We therefore will not assume any specific form of the distribution of the  elements of $H$-matrix except that the latter must  satisfy basic symmetry and conservation laws of the Hamiltonian.

The paper is organized as follows.  Section II describes the Hamiltonian of a basic block as a collection of molecules and calculates the variance of its matrix elements in a physically motivated basis. The latter information is used in section III to derive its ensemble averaged density of states which turns out to be a semicircle in the bulk of the spectrum but growing super-exponentially in its edge. As the density is derived without using any system-specific information about the intra-molecular forces, it is applicable for  a wide range of glasses. Section IV derives the specific heat of the basic block for the temperature ranges below and above Debye temperature and confirms its linear/superlinear temperature dependence even at nanoscale sizes. We conclude in section V with a summary of our main ideas, results and a brief comparison with some existing theories. As the theoretical formulation discussed here requires a. State of the art knowledge of many interdisciplinary research areas e.g. glass chemistry (for microscopic structure), glass physics (for interactions), system dependent random matrix theory, standard statistical physics as well as mathematical Airy functions, it is practically not feasible to review all of them in the main text. Some technical derivations are however illustrated in the appendices, with  others deferred to the supplemental material \cite{sup} to avoid defocussing of the paper.

\section{The Hamiltonian}

An appropriate description of the Hamiltonian of the basic block requires a prior knowledge of the microscopic structure, local order and interactions in  amorphous materials. For completeness sake, we briefly review it here and also in supplementary material \cite{sup}.

\subsection{Hierarchy of order and local interactions}

Based on extensive experimental evidence ( see e.g. \cite{ell1,sbb,cheng} and references therein), the structural order in amorphous materials is classified in two length scales:

 (i) {\it Short range order (SRO)}:  associated with well-defined first neighbors arrangements and  characterized in terms of bond lengths and bond angle distributions, the structure within SRO length scales (hereafter referred as SRO structure) are very stable and often similar to crystalline material (the one lying nearest in composition) \cite{ell1}. As experiments indicate,  the intermolecular interaction between two SRO structures is much weaker than the interaction within a SRO (later referred as intra-SRO forces) \cite{phil2,b2008,b1991,b1999}. For example, the SRO structure in three dimensional network of a $\beta$-crystoballite glass consists of $Si[SiO_4]$ units (see page 39 of \cite{b1994}). Here two cations  (Si atoms)  are completely screened by four anions  ($O$-atoms) thus ruling out any strong binding (electrostatic)  interaction between two neighboring SRO structures  \cite{ell1,cheng}. 
 
(ii) {\it Medium range order (MRO)}: 
Although an exact understanding of MRO structure remains elusive even now \cite{sbb, cheng}, it can in general be regarded as the next highest level of structural organisation beyond SRO \cite{ell1}. For example, MRO in covalent glasses typically refers to structural features in the range of $5 \to 20\; \AA$, including varying arrangement of connected polyhedra e.g. rings, chains or layers, superstructural units and clusters e.g. those around network modifiers \cite{sbb}.  Some more details of SRO and MRO structures including the example  of $\beta$-cristobalite glass, taken from studies \cite{b2008,b1991,b1999}, are discussed in section I of the supplementary part \cite{sup}.

The hierarchy of length scales in turn is associated with a hierarchy of local interactions.  The constraints theory of glass structure \cite{phil2} suggests a classification of intermolecular forces into strong and weak types, with former acting as the mechanical constraints.  The stability of glass structure requires the number of constraints per atom exactly equal the number of degree of freedom. Experiments however reveal broadening of second nearest neighbor peak  in diffraction patterns; the configurational strain energy required for  the broadening  can only be contributed by the residual Vanderwaal interactions \cite{phil2}. A typically higher strength of forces at SRO  than those at MRO, is also indicated by the experimentally observed high stability of SRO structures  and deformation and randomness associated with those at MRO \cite{sbb,cheng}.  While the intra-SRO forces are typically pure binding e.g. covalent or ionic type, those connecting SROs  are usually mixed binding forces, subjected to deformation and polarization and thereby resulting in  Vanderwaal forces \cite{sme, phil2}.  The presence of network modifiers e.g. $Na_2O$ in oxide glasses also gives rise to  non-bridging oxygens appearing as dangling bonds; the lone pair of electrons in these bonds again leads to Vanderwaal interactions \cite{b1991, b2008}. In fact mixed bonding is now believed to be essential for glass formation and purely covalent bonding  to induce crystalline order \cite{b2008, b1991,b1999,b1994}.

\subsection{Intermolecular interactions among SRO structures}

 With both crystalline and non-crystalline materials sharing similar structure at SRO and their differences emerging at MRO scales \cite{phil2,ell1,b2008}, intuitively the latter seems to be the most potential candidate to explain their differences. Many previous studies have also suggested the role of MRO structures in the peculiarities of the vibrational spectrum e.g. boson peak (e.g. see table 1.2 of \cite{b1999}). This motivates us to analyze  the energy level statistics of a block of MRO length scale; its Hamiltonian can be expressed as a sum over the  Hamiltonians of free SRO structure subjected to intermolecular interactions.  (This is equivalent to consider the  block structure arising when the free SRO molecules initially far apart, brought together at  non-overlapping distances ranging between SRO and MRO scales).

With SRO structures being highly stable by definition, the interactions among any two of them is much weaker and is typically of Vanderwaal type (at non-overlapping distances).
For non-polar SRO molecules, the dominant contribution to inter-SRO interactions arise from quantum induced instantaneous  dipole-induced dipole interactions (referred as dispersion forces in short) and caused by a correlated movements of electrons in interacting molecules. In an attempt to avoid each other at short intermolecular distances, a redistribution of  electrons that belong to different molecules leads to formation of instantaneous dipoles that attract each other \cite{ajs}.  In case of molecules with permanent dipole moments, other types of VW forces appear too i.e Debye forces (permanent dipole-induced dipole interaction, also referred as induction forces) and Keesom forces (interactions among permanent dipoles that are ensemble averaged over different  rotational orientations of the dipoles). The dispersion force however contribute the most to overall inter-molecule bonding even in presence of other electrostatic interactions e.g. hydrogen bonding or Keesom interaction \cite{ajs}. 

As indicated by extensive experimental evidence  in past \cite{tang1}, the sum of energies of the separated atoms is  only slightly different from the energy of the interacting atoms. Although at large distances i.e zero overlap of their charge distribution, the polarisation  interactions become significant but their relative weakness, compared to sum of the energies of free atoms, permits a second-order perturbation treatment. This motivated many models of intermolecular potential e.g. Tang and Tony (TT)  model \cite{tang1}, BKS model \cite{bks}, TTAM model \cite{ttam} etc.  Based on these models, the interaction potential for a pair of atoms with charges $Z_1, Z_2$ at a distance $R$ can be given  as \cite{tang1,bks}: ${\mathcal U}(R)={\mathcal U}^{coul}+ {\mathcal U}^{rep}(R) + {\mathcal U}_{pol}(R)$.
Here ${\mathcal U}^{coul}$ is the coulomb interaction between the atomic charges (for example, BKS model gives  ${\mathcal U}^{coul} = {Z_1 Z_2 \over R}$) and ${\mathcal U}_{rep}$ is a generalized  Born-Mayer repulsive potential: $U_{rep}(R)=A \; {\rm exp}(-s R)$ with $A$ as  the effective  amplitude  for the repulsive potential  at short distances  \cite{bks} and $s$ as a measure of the screening of the nuclear charge by electrons, ${\mathcal U}_{pol}(R)$ is the polarisation contribution in a damped series form: ${\mathcal U}_{pol}(R) = - \sum_{n \le 3}^{\infty} f_{2n}(s R) \; {C_{2n} \over R^{2n}}$ with $C_{2n}$ as dispersion coefficients, determined by abinitio routes,  and $f_{2n}(b_c R) $ as the damping coefficeints: $f_{2n}(x) =1- {\rm e}^{-x} \sum_{k=0}^{2n} {x^k \over k!}$.

In case of a cluster of SRO molecules, the pair-potential mentioned above  can be generalized to account for the many body effects \cite{mad,ajs}. Consider a block composed of $g_0$ such molecules, each of mass $m$ and  labeled as $''qk'' $ with its centre at position ${\bf r}_k$, $k=1 \to g_0$, in an arbitrary  amorphous material block of MRO length scale.
The Hamiltonian ${\mathcal H}$ of the block  can then be written as a sum over the Hamiltonians of the separated (uncoupled or free) molecules and the potential energy acquired in bringing them together at their positions ${\bf r}_k$,
\begin{eqnarray}
{\mathcal H} ={\mathcal H}_0 + {\mathcal U}({\bf r}_1, {\bf r}_2,\ldots, {\bf r}_{g_0})
\label{hb1}
\end{eqnarray}
where ${\mathcal H}_0$ is the total Hamiltonian of $g_0$ uncoupled (or free) molecules
\begin{eqnarray}
{\mathcal H}_0 &=& \sum_{n=1}^{g_0} \;   h_n({\bf r}_n).  
\label{hb10} 
\end{eqnarray}
Here $ h_n({\bf r}_n)  $ is the Hamiltonian of the $n^{th}$ SRO structure (hereafter referred as ''molecule'' for brevity unless necessary).  Note as a SRO molecule is held by strong pure binding forces, $h_n({\bf r}_n)$ includes the covalent/ionic bonding energy too.

The ${\mathcal U}$ is the sum over inter-molecular interactions of SRO molecules lying within MRO length scales. The inter-SRO interactions being relatively weaker as compared to intra-SRO ones \cite{phil,sme,ell1} (subjected only to mixed bonding \cite{sme,phil2,b2008, b1994, b1999}, ${\mathcal U}$ can be described as a sum over  coulomb and repulsive potential energies and  a perturbation series of polarization corrections only \cite{mad,ajs}:
\begin{eqnarray}
{\mathcal U}\left({\bf r}_{1}, \ldots, {\bf r}_{g_0}\right) ={\mathcal U}^{coul}+ {\mathcal U}^{rep} + {\mathcal U}^{pol}.
\label{u1}
\end{eqnarray} 
Here ${\mathcal U}^{coul}\left({\bf r}_{1}, \ldots, {\bf r}_{g_0}\right)$ and ${\mathcal U}^{rep} \left({\bf r}_{1}, \ldots, {\bf r}_{g_0}\right)$  are the potential energies due to coulomb  and repulsive interactions, respectively, among  SRO molecules and 
and ${\mathcal U}^{pol}\left({\bf r}_{1}, \ldots, {\bf r}_{g_0}\right)$ refers to  the polarization interactions (i.e induction and dispersion) among $g_0$ molecules. 

 The coulomb energy  ${\mathcal U}^{coul}$ can be written in a generic form as  
\begin{eqnarray}
{\mathcal U}^{coul} \left({\bf r}_{1}, \ldots, {\bf r}_{g_0}\right) = \sum_{n,m; \atop n \not=m} {Q ({\bf r}_n, {\bf r}_m) \over |{\bf r}_n -{\bf r}_m|}
\label{ucl0}
\end{eqnarray}
 with $Q ({\bf r}_n, {\bf r}_m)$ as a meausre of the screened charge distributions on the molecules at ${\bf r}_n$ and ${\bf r}_m$ \cite{bks,ttam,mad}. For example, in BKS model, $Q=Z_m Z_n$ with $Z_m, Z_n$ as the molecular charges \cite{bks}. 
 
As ${\mathcal U}^{rep}$ arises due to  the repulsion at small separations, it is  essentially a pairwise potential and can be expressed as a sum over pairwise generalised Born-Mayer potentials: 
\begin{eqnarray}
{\mathcal U}^{rep}\left({\bf r}_{1}, \ldots, {\bf r}_{g_0}\right) = \sum_{a,b\atop a \not=b}  A^{(a,b)} \; {\rm exp}(-s |{\bf r}_a-{\bf r}_b|)
\label{urep}
\end{eqnarray}
with $A^{(a,b)}(|{\bf r}_a-{\bf r}_b|)$ as an operator determining the amplitude of the repulsive potential between the molecules at positions ${\bf r}_a$ and ${\bf r}_b$. Contrary to the case of a single molecule-pair mentioned above, $A^{(a,b)}$, in a cluster of molecules is complicated and  depends on the relative orientation of the pair (e.g. see eq.(3) of \cite{mad}). Its detailed form however is not needed for our analysis.

The energy ${\mathcal U}^{pol}$ can be written as a summation over $n$-body contributions, referred as ${\mathcal U}_{pol}^{(q1,q2,..,qn)}$ (arising from polarization interactions among $n$ molecules, labeled as $''q1'', ''q2''...,''qn''$, $n=2 \to g_0$):

\begin{eqnarray}
{\mathcal U}^{pol} \left({\bf r}_{1}, \ldots, {\bf r}_{g_0}\right) =\sum_{n=2}^{g_0}  \; \sum_{q1,q2,\ldots, qn} {\mathcal U}_{pol}^{(q1,q2,\ldots, qn)} \left({\bf r}_{q1}, \ldots, {\bf r}_{qn}\right)
\label{upol}
\end{eqnarray}
Here the symbol $\sum_{q1, q2,..,qn}$ corresponds to a summation  over all possible n-body polarization interactions  (i.e including both dispersion as well as induction type) among all $g_0$ molecules.

\subsection{Matrix Representation}

The eigenvalues of an operator are basis independent. Consequently, to analyze any property that is a function of the eigenvalues only, the operator can be matrix represented in an arbitrary basis. To analyze the  specific heat behavior, a function of the energy levels, here we consider the matrix representation of ${\mathcal H}$  in the eigenbasis of the non-interacting SRO  molecules (hereafter referred as NIM basis). The latter consists of the direct-product of  $g_0$ normalised single SRO molecule states  in which the molecules are assigned to definite single-molecule states. 
 
  In general a single SRO molecule can exist in  infinitely many states of  electronic, vibrational and rotational types; this would imply an infinite dimensional basis-space. To study the thermal effects at very low temperatures ($T < 30 K$), it is however sufficient to consider only rotovibrational levels in the electronic ground state of each isolated molecule \cite{exp1}. 
Assuming  only ${\mathcal N}$ such states of each molecule participate in the interaction, this gives the size of NIM basis space as 
\begin{eqnarray}
N = {\mathcal N}^{g_0}.
\label{n1}
\end{eqnarray}
 Let  $| {\mathcal K}_n \rangle$ and $E_{ {\mathcal K}_n} =\langle {\mathcal K}_n| h_n |{\mathcal K}_n\rangle$, with $ {\mathcal K}_n=1,2,...{\mathcal N}$ be the eigenvectors and eigenvalues of a molecule, say "$n$" with its Hamiltonian as $h_n$.
A typical basis-state in the $N \times N$ dimensional NIM basis can then be given as 
\begin{eqnarray}
|{\mathcal K} \rangle  = \; \prod_{n=1}^{g_0} \; |{ {\mathcal K}_n}\rangle
\label{nim1}
\end{eqnarray}
 with  $|{ {\mathcal K}_n}\rangle$ referring to  the specific eigenvector of the $n^{\rm th}$ molecule which occurs in the product state $|{\mathcal K}\rangle$. (Note as the exchange interactions are neglected,  the basis need not be anti-symmetrized).    
From eq.(\ref{hb10}),  ${\mathcal H}_0$ is a diagonal matrix in the NIM basis 
\begin{eqnarray}
{\mathcal H}_{0; {\mathcal KL}}   \equiv  \langle {\mathcal K} |{\mathcal H}_0| {\mathcal L} \rangle 
= \sum_{n=1}^{g_0} \; E_{ {\mathcal K}_n} \; \delta_{{\mathcal KL}}
\label{h1kl}
\end{eqnarray}

Eq.(\ref{u1})  leads to the element ${\mathcal U}_{\mathcal KL} \equiv \langle {\mathcal K} |{\mathcal U}| {\mathcal L} \rangle $ corresponding to interaction energy for transition from state ${\mathcal K}$ to ${\mathcal L}$. 
Due to ${\mathcal U}_{pol}^{(q1,q2,..,qn)}$ being a $n$-body interaction among the molecules, the NIM basis has selection rules associated with a $n$-body interaction \cite{ajs}; only $n$ molecules labeled as $''q1, q2, \ldots, qn''$ can be transferred by $H$ to different single-molecule states; Its matrix element  between basis states $| {\mathcal K} \rangle$ to $|{\mathcal L} \rangle$ can then be given as  
\begin{eqnarray}
\langle {\mathcal K} |{\mathcal U}_{pol}^{(q1,q2,..,qn)}| {\mathcal L} \rangle 
\; \; =  \; \;  D^{(q1,q2,..qn)}_{\mathcal K \mathcal L} \;\;  {\mathcal F}^{(q1,q2,..,qn)}_{\mathcal KL}
 \label{u1klt}
\end{eqnarray}
where  $ {\mathcal F}^{(q1,q2,..,qn)}_{\mathcal KL} \left({\bf r}_{q1}, \ldots, {\bf r}_{qn}\right)$ is a function dependent on polarisability of the molecules $''q1'', ''q2'',..''qn''$ as well as their separation (its generic form very complicated and  discussed in \cite{ajs}) and 
\begin{eqnarray}
D^{(q1,q2,..qn)}_{\mathcal K \mathcal L} = \prod_{q=1 \atop q \not=q1,q2,..,qn}^{g_0} \; \delta_{ {\mathcal K}_q, {\mathcal L}_q}  \;  
\label{d1kl}
\end{eqnarray}
As clear from the above, the matrix element in eq.(\ref{u1klt}) is non-zero only if the   basis-pair $|{\mathcal K}\rangle, |{\mathcal L} \rangle $ have same contributions from the rest $g_0-n$ molecules  (excluding those labeled by "$q1,q2,..,qn$"); we henceforth refer a pair $|{\mathcal K}\rangle, |{\mathcal L} \rangle $ as $p$-plet if they differ in eigenfunction contributions from $p$-molecules i.e ${\mathcal K}_m \not={\mathcal L}_m$ with $m$  taking $p$-values from the set  $\{1,2, \ldots, g_0\}$. Clearly the total number of states $|{\mathcal L}\rangle$ forming a $p$-plet with a fixed $|{\mathcal K}\rangle$ is $(1/2) g_0! (g_0-1)! ({\mathcal N}-1)^p$.

As the induction and dispersion interactions result in transition of  molecules from one state to another, they  contribute only to off-diagonal matrix elements (as discussed in chapter 4 of \cite{ajs}). From eq.(\ref{u1}), the matrix element ${\mathcal H}_{\mathcal KL} $ can then be written as ${\mathcal H}_{0; {\mathcal KL}} + {\mathcal U}_{\mathcal KL} $ where
\begin{eqnarray}
{\mathcal U}_{\mathcal KL} 
= \delta_{\mathcal KL} \;  {\mathcal U}^{coul}_{\mathcal KL} + \sum_{a,b \atop a \not=b}  A^{(a,b)}_{\mathcal KL}   \; {\rm exp}(-s |{\bf r}_a-{\bf r}_b|)   + \left(1-\delta_{\mathcal KL} \right) \; 
\sum_{n=2}^{g_0}  \; \sum_{q1,q2,\ldots, qn} D^{(q1,q2,..qn)}_{\mathcal K \mathcal L}  \; \; {\mathcal F}^{(q1,q2,..,qn)}_{\mathcal KL} \nonumber \\
\label{u2klt}
\end{eqnarray}
Note as the coulomb interaction is relatively small at non-overlapping distances of the two molecules, it only contributes to diagonals (as also evident from the BKS coulomb potential \cite{bks}). 

For  ${\mathcal K, L}$ pair forming a $p$-plet (for $p >0$), ${\mathcal U_{\mathcal KL}}$ has contribution  from all $p+x$ body induction and polarization terms with $x \ge 0$ (i.e $p \le p+x \le g_0$) (see supplementary part \cite{sup}).  Although an exact calculation of these matrix elements is possible only for simple molecules,  they can be approximated by some well-established methods e.g. using model potentials or by a multi-pole expansion of the $n$-body operator; the coefficients of expansions are then determined by abinitio approaches (see e.g.\cite{bks,ttam,ajs}).

The matrix elements given by eq.(\ref{h1kl}) and eq.(\ref{u2klt}) are valid for both polar as well non-polar molecules. But to gain better insight, without any loss of generality, henceforth we  focus on non-polar molecules only; the polarization contribution  in that case consists of dispersion interaction only, thus simplifying the series in eq.(\ref{u2klt}). Contrary to induction,  a $n$-body dispersion interaction corresponds to $n$ instantaneous dipoles  interacting with each other and results in change of the states of all $n$-molecules. Its  contribution to a matrix element in the NIM basis  is therefore non-zero only if the pair of basis-states form an $n$-plet. 
For sufficiently large distances, say $R$ between  molecule pairs ($R > 10 \; a_0$ with $a_0$ as Bohr radius),  the $n$-body term can  be well-approximated by a multi-pole expansion in which the leading order term is  of type $R^{-3n}$ (arising from $n$-dipoles interaction) \cite{ajs}. For qualitative estimates of the energy, it is sufficient to take the first term in the expansion and one can write \cite{ajs, mad} 
\begin{eqnarray}
{\mathcal F}^{(q1,q2,..,qn)}_{\mathcal KL} 
\approx   { C^{(3n)}_{\mathcal KL} \over |{\bf r}_{q1}-{\bf r}_{q2}|^{3} ...|{\bf r}_{qn}-{\bf r}_{q1}|^{3}  }.
\label{fkl}
\end{eqnarray}
with $C^{(3n)}_{\mathcal KL} $ as the coefficient of term corresponding to $n$-instantaneous dipoles interaction in multi-pole expansion of ${\mathcal U}_{\mathcal KL}$. Substitution of the above in eq.(\ref{u2klt}) and using the latter  along with eq.(\ref{h1kl}) in eq.(\ref{hb1}), one can write 
\begin{eqnarray}
{\mathcal H}_{{\mathcal KL}}  &\approx & \delta_{{\mathcal KL}} \;\left(\sum_{n=1}^{g_0} \; E_{ {\mathcal K}_n} + {\mathcal U}^{coul}_{\mathcal KK}\right)+ \sum_{a,b \atop a \not=b}  A^{(a,b)}_{\mathcal KL}  \; {\rm e}^{-s |{\bf r}_a-{\bf r}_b|} + 
 { (1-\delta_{{\mathcal KL}}) \; C^{(3n)}_{\mathcal KL} \over |{\bf r}_{q1}-{\bf r}_{q2}|^{3} ...|{\bf r}_{qn}-{\bf r}_{q1}|^{3}} 
\label{h2kl}
\end{eqnarray}

Eq.(\ref{ucl0}) gives  ${\mathcal U}^{coul}_{\mathcal KK} = \sum_{n,m; \atop n \not=m} {Q_{\mathcal KK} ({\bf r}_n, {\bf r}_m) \over |{\bf r}_n -{\bf r}_m|}$.  As the dominant contribution comes here from the  nearest neighbor molecules, it can be simplified. Assuming $z$ as the number of nearest neighbors of a SRO molecule and  $2 R_v$ as the average distance of closest approach between them, this gives 
\begin{eqnarray}
{\mathcal U}^{coul}_{\mathcal KK} = {z \; g_0  \over 2 R_v} \; Q_{\mathcal KK}.
\label{ucl}
\end{eqnarray}

The $N \times N$ matrix ${\mathcal H}$ in the NIM basis  has total $N(N+1)/2$ elements given by eq.(\ref{h2kl}).  Although  the matrix elements corresponding to $p$-plets, with $p$ large,  are relatively weaker but  their  total number compensates for the weakness. This implies ${\mathcal H}$  effectively behaves as a dense matrix, with total contribution from the offdiagonals much larger than the diagonals.

\subsection{Statistics of the Matrix Elements}

As mentioned in the beginning of section II.C, our objective is to calculate the heat capacity which in turn requires a knowledge of the eigenvalues of the Hamiltonian and therefore its matrix elements in any basis. In case of  a complex system such as a typical amorphous system, it is often difficult to exactly determine the matrix elements and they often fluctuate from one sample to the other. This in turn affects the eigenvalues and thereby all the physical properties based on them and makes it necessary to consider their statistical behavior.  In the domain of complex systems studies, therefore, it is an unavoidable standard requirement to know a priori the statistics of the matrix elements (i.e how the latter are distributed over an ensemble of the exact replicas of the system Hamiltonian). Here we consider the statistics of the  ${\mathcal H}_{{\mathcal KL}}$ given by eq.(\ref{h2kl}).

From eq.(\ref{h1kl}), the diagonals of ${\mathcal H}_0$ in the NIM basis are a sum over single molecule energy levels. As SRO molecules are typically polyatomic, the couplings between various degrees of freedom (e.g electronic, vibrational, nuclear)  of a single molecule can not be ignored even for low lying energy states \cite{zkpcd}. 
This in turn renders the description of each energy level by  a set of quantum numbers often very difficult and meaningless, leaving a statistical analysis of the spectrum as the only option. The basic prerequisite of such an analysis, i.e the availability of a sufficiently large data set of energy levels of a single molecule is again experimentally not feasible \cite{zkpcd}.  This has  motivated in past consideration of the averaging  over an ensemble of the exact replicas of a molecule \cite{zkpcd} (also see {\it appendix A} and section 4.2.2 of \cite{gmw}). Assuming ergodicity, the energy level fluctuations along the spectrum of a molecule are then assumed to be given by those over the ensemble.

The coefficients $C^{(3n)}_{\mathcal KL}$, referring to the strength 
of $n$-instantaneous dipole interactions, as well as the amplitudes ${\mathcal A}^{(a,b)}_{\mathcal KL}$   depend on the mutual orientation of the SRO molecules within a block of MRO length scale (e.g. see eqs.(3,5) of \cite{mad} and also \cite{ajs}). In an ensemble of replicas of such blocks, however, the orientations are known to vary rapidly from one replica to the other, resulting in both positive as well negative contributions from ${\mathcal F}^{(q1,q2,..,qn)}_{\mathcal KL}$,  even if one considers the same set of $n$-molecules in each replica block. Further all  mutual orientations of the molecules are equally probable if one considers a sufficiently large ensemble. Consequently both $C^{(3n)}_{\mathcal KL}$ and  ${\mathcal A}^{(a,b)}_{\mathcal KL}$ can be assumed to behave as  random variables, with a finite variance, over an   ensemble of the basic block replicas and their averages over  a sufficiently large ensemble, for a fixed ${\mathcal K,L}$-pair states, can  safely be  assumed to be zero i.e $\langle C^{(3n)}_{\mathcal KL} \rangle=0$, $\langle {\mathcal A}^{(a,b)}_{\mathcal KL} \rangle=0$.

The above in turn leads to a randomization of ${\mathcal H}_{kl}$ with its distribution parameters determined as follows. (As indicated by the experiments, the location of a specific SRO molecule, say ${\bf r}_n$, can also vary from one replica to other but is relatively negligible and is not considered here). An ensemble average of eq.(\ref{h2kl}),  followed by substitution of $\langle C^{(3n)}_{\mathcal KL}\rangle=0$ and $\langle {\mathcal A}^{(a,b)}_{\mathcal KL} \rangle=0$ gives 
\begin{eqnarray}
 \langle {\mathcal H}_{\mathcal KL} \rangle  &\approx&  \delta_{KL} \; \left( \sum_{n=1}^{g_0} \langle E_{{\mathcal K}_n} \rangle + \langle {\mathcal U}^{coul}_{\mathcal KK} \rangle \right) 
  \label{hav1}
 \end{eqnarray} 
 with $\langle . \rangle$  implying an average over an ensemble of the exact basic block replicas.

As mentioned above, the allowed number of dipole transitions permit only three vibrational states of each molecule to be involved; without loss of generality, these can be referred as ${\varepsilon}-\Delta {\varepsilon}, {\varepsilon}, {\varepsilon}+\Delta {\varepsilon}$ (with typical values of ${\varepsilon}, \Delta {\varepsilon} \sim 10^{-20} \; J$). With each $E_{{\mathcal K}_n}$ allowed to take any of the three values, clearly the spectral average ${\overline{E_{{\mathcal K}_n}}} =\varepsilon$. Assuming the ensemble of single molecule replicas to be ergodic, the spectral average of $E_{{\mathcal K}_n}$ can then be replaced by its ensemble  average; this leads to  $\langle E_{{\mathcal K}_n}\rangle = {\overline{E_{{\mathcal K}_n}}} =\varepsilon$. Substituting the latter along with eq.(\ref{ucl})  in eq.(\ref{hav1}), we have
\begin{eqnarray} 
 \langle {\mathcal H}_{\mathcal KK} \rangle = g_0 \; \varepsilon_0.
\label{mean}
\end{eqnarray} 
where $\varepsilon_0 =\varepsilon+  {z \; g_0 \over 2 R_v} \; \langle Q_{\mathcal KK} \rangle$. Further the stability of a SRO structure is usually attributed to a complete screening of  cations by surrounding anions, thus ruling out any strong  electrostatic  interaction between two neighboring SRO structure. For example, for $SiO_4$ tetrahedra, \cite{bks} gives  $q_O=-1.2$ and $q_{si}=2.4$ thus implying a zero net charge on the SRO molecule $Si[SiO_4]$; the Coulomb interaction between two SRO molecules is therefore  almost negligible i.e $\langle Q_{\mathcal KK} \rangle \sim 0$. (Even for the cases, where latter is not a good approximation, it would only change the origin of the spectrum).

Squaring eq.(\ref{h2kl}) followed by its ensemble averaging gives the second moment $v_{\mathcal KL} = \langle ({\mathcal H}_{\mathcal KL})^2 \rangle $.
As $E_{\mathcal K_m}$ refers to the  energy level of a free molecules, the correlations between energy levels of two different free molecules can be neglected: $\langle  E_{\mathcal K_m}  E_{\mathcal K_n} \rangle  \approx \delta_{mn}$. Similarly the terms with many body correlations  $\langle {\mathcal A}^{(a,b)}_{\mathcal KL} C^{(3n)}_{\mathcal KL} \rangle \approx \langle {\mathcal A}^{(a,b)}_{\mathcal KL} \rangle .\langle C^{(3n)}_{\mathcal KL} \rangle  \approx 0$  and can be ignored. This leads to
\begin{eqnarray}
 v_{\mathcal KL}    &\approx & \delta_{\mathcal KL} \; g_0 \; \nu_0 +z \; g_0 \; \langle ({\mathcal A}^{(neigh)}_{\mathcal KL} )^2 \rangle \; {\rm e}^{-4 s R_v} + {(1-\delta_{\mathcal KL}) \; \langle (C^{(3n)}_{\mathcal KL})^2 \rangle \over \langle|{\bf r}_{q1}-{\bf r}_{q2}|^{6} ...|{\bf r}_{qn}-{\bf r}_{q1}|^{6} \rangle }  
\label{vnp}
\end{eqnarray}
with $\nu_0=\langle  E_{\mathcal K_n}^2\rangle+{z^2\over 4 R_v^2} \; \langle Q_{\mathcal KK}^2\rangle + {z\over  R_v} \langle  E_{\mathcal K_n} \rangle \langle Q_{\mathcal KK}\rangle $. The $2nd$ term in eq.(\ref{vnp}) is obtained as follows:  squaring of eq.(\ref{h2kl}) leads to a term $S_1 \equiv \sum_{a,b, a',b'} \langle {\mathcal A}^{(a,b)}_{\mathcal KL}  {\mathcal A}^{(a',b')}_{\mathcal KL} \rangle \;  \langle{\rm e}^{-s |{\bf r}_a -{\bf r}_b| -s |{\bf r}_{a'} -{\bf r}_{b'}|} \rangle$. Neglecting the relatively weaker four body correlations  $\langle {\mathcal A}^{(a,b)}_{\mathcal KL} {\mathcal A}^{(a',b')}_{\mathcal KL} \rangle$ gives $S_1 \approx \sum_{a,b} \langle \left({\mathcal A}^{(a,b)}_{\mathcal KL} \right)^2\rangle \;  \langle{\rm e}^{-2 s |{\bf r}_a -{\bf r}_b|}\rangle$. Here again the dominant contribution comes from the  nearest neighbor molecules resulting in  $S_1 \approx  z \; g_0 \; \langle ({\mathcal A}^{(neigh)}_{\mathcal KL} )^2 \rangle \; {\rm e}^{-2 s R_v}$ with $\langle \left({\mathcal A}^{(a,b)}_{\mathcal KL} \right)^2\rangle$   approximated  as $\langle ({\mathcal A}^{(neigh)}_{\mathcal KL})^2 \rangle$ for all neighboring pairs.

The  correlations among different matrix elements can be assumed to be  negligible e.g  $\langle {\mathcal H}_ {\mathcal KL} \; {\mathcal H}_{mn} \rangle_e \approx 0  $ if $(k, l) \not= (m,n)$; this is again  due to presence of both positive as well as negative contributions which on summation cancel each other. Proceeding similarly and using same approximations, higher order moments of ${\mathcal H}_{\mathcal KL}$ can also be calculated.

\subsection{ Bulk-Spectral Parameter $b$}

 As  discussed in next section, a determination of the density of the states of the basic block requires a knowledge of the bulk spectral parameter defined as $b =\left(2\sum_{\mathcal L} v_{\mathcal KL}\right)^{-1/2}$. It can be determined as follows.
 
 With ${\mathcal K, L}$ referring to the vibrational states of the  electronic ground state, the coefficients  $C^{(3n)}_{\mathcal KL}$ are expected to be of the same order for all such basis-pairs and can be approximated by their average value over 
all states. Let us define ${\mathcal C}_{3n}^2$ as the  mean square of the coefficient of the $n$-dipole interaction, respectively,  with averaging over all the states as well over the ensemble (equivalent to averaging over all orientations):
\begin{eqnarray}
{\mathcal  C}_{3n}^2 &=& {1\over N^2}\sum_{\mathcal K, L} \; \langle (C^{(3n)}_{\mathcal KL})^2 \rangle. 
\label{cav}
\end{eqnarray}
 Approximating the contributions from all single molecule vibrational states undergoing dipole transition  as almost same, one can write
\begin{eqnarray}
\sum_{\mathcal L=1}^{N}  v_{\mathcal KL} \approx   g_0 \; \nu_0 + z \; g_0 \; A_{rep}^2 \; {\rm e}^{-4 s R_v}
+  \sum_{n=2}^{g_0}   \; \eta^n \;  {\mathcal C}_{3n}^2  \; \sum_{q1,..,qn =1 \atop q1 \not= q2 ..\not=qn}^{g_0} {1\over |{\bf r}_{q1}-{\bf r}_{q2}|^{6} ...|{\bf r}_{qn}-{\bf r}_{q1}|^{6}  }.
\label{vnps0}
\end{eqnarray}
with $A_{rep}^2 = \sum_{\mathcal L=1}^{N} \langle ({\mathcal A}^{(neigh)}_{\mathcal KL} )^2 \rangle $ and
 $\eta= {\mathcal N}-1$ as the number of allowed dipole transitions among the ${\mathcal N}$ eigenstates of  a single molecule, say  $''q_n''$ from a fixed state ${\mathcal K}_{q_n}$ \cite{note1}.  
 The third term on the right of eq.(\ref{vnps0}) can further be simplified as follows: as $n$- body dipole-dipole interaction
contains a factor of $r^{-3n}$, four body and higher many body terms are expected to be
less important. Besides, no more than four atoms can be in mutual contact (see page 14, section 10.2 of \cite{ajs}). 
For order of magnitude calculations, therefore,  it suffices to keep the 
contribution from the nearest neighbor SRO molecules (referred by $z$) only.
Also note that the total number of states ${\mathcal L}$ forming a $n$-plet (defined above eq.(\ref{u2klt})) with a given state ${\mathcal K}$ and involving transitions of nearest neighbor molecules only is $z \; {\mathcal N}^n$. 
This leads to  
\begin{eqnarray}
\sum_{\mathcal L=1}^{N}  v_{\mathcal KL} \approx  g_0 \; \nu_0 + z \; g_0 \; A_{rep}^2 \; {\rm e}^{-4 s R_v}
+ {z \; g_0 \; \eta^2  \; {\mathcal C}_{6}^2\over 2 \; (2 R_v)^{12}}  + {
z(z-1) \; g_0 \; \eta^3   \; {\mathcal C}_{9}^2 \over 4 \; (2 R_v)^{18}}  
\label{vnps}
\end{eqnarray}

As the contribution from Coulomb terms is either negligible or is, at best, of the order of $\langle  E_{\mathcal K_n}^2\rangle$, $\nu_0$, defined below eq.(\ref{vnp}), can be estimated as $\sim\langle  E_{\mathcal K_n}^2\rangle$. Further note $E_{{\mathcal K}_n}$, the energy level of a SRO molecule,  itself is a sum over contributions from  the constituent atoms as well as their interaction energies and $\langle  E_{\mathcal K_n}^2\rangle-\langle  E_{\mathcal K_n}\rangle^2$ corresponds to the square of the natural line-width of a typical vibrational level due to Heisenberg uncertainly and is negligible at very low temperatures. Thus taking  the natural line width of a typical vibrational level $\sim 10^{-11} \; cm^{-1}$ ($\approx 10^{-34} \; J$) gives $\nu_0 \sim (10^{-34})^2 \; J^2$.

Detailed studies of the higher order dispersion coefficient $C_{3n}$ for various molecules indicate their rapid decay with increasing $n$  (see table 10.1 and section 10.2 of \cite{ajs}) and typically  ${\mathcal C}_6 \sim 10^2 \; eV-\AA^6$ and ${\mathcal C}_9 \sim 10^{-8} \; eV-\AA^9$. More specifically, for silica, \cite{bks}  gives $A_{rep} \sim  10^4\; {eV},s \sim 2.76 \; \AA^{-1}, C_6 \sim 175.0 \; eV-\AA^6$ for $Si-O$ bond and $A_{rep} \sim 10^3 \; {eV}, s \sim 4.87 \; \AA^{-1}, C_6 \sim 133.54 \; eV-\AA^6$ for $O-O$ bonds. With $R_v \sim 3-5 \; \AA$ for amorphous material (\cite{bb3}), the contributions from the terms containing $\nu_0 \; (\approx 10^{-30} \; eV^2)$, $A_{rep}$ as well as ${\mathcal C}_9$ are then negligible as compared to the ${\mathcal C}_6$-term \cite{cn, more}. Eq.(\ref{vnps} can now be approximated as 
\begin{eqnarray}
 {1\over 2 \; b^2} = \sum_{\mathcal L=1}^{N}  v_{\mathcal KL}  \approx  { z \; g_0  \; \eta^2 \;  {\mathcal C}_{6}^2  \over 2  \; (2 R_v)^{12}}.   
\label{vnpsa}
\end{eqnarray}

The coefficient $C_6$ can further be written in terms of the Hameker constant $A_H$ (a constant for materials) i.e $A_H \approx \pi^2 \; C_6 \; \rho_n^2$, with $\rho_n$ as the number density of the molecules.  
Taking $\Omega_{\rm eff} = {1\over \rho_n}$ as the average volume available to a typical glass molecule, we have 
\begin{eqnarray}
\Omega_{\rm eff} = s_m \; (R_v +R_m)^3 \approx  (1+y)^3 \; \Omega_m
\label{ome}
\end{eqnarray}
with $\Omega_m$  the molar volume: $\Omega_m = s_m \;  R_m^3$, with $s_m$ as a structure constant  e.g. $s_m =4 \pi/3$ assuming a spherical shape for the molecule.  Here $R_v$ is half the distance between two nearest neighbor molecules;  the 2nd equality in eq.(\ref{ome}) follows by writing $R_v = y \; R_m$. The above in turn gives
\begin{eqnarray}
C_6 \approx {A_H \; (\Omega_{\rm eff})^2 \over \pi^2} \approx {{ s_m^2 \over \pi^2} \;(1+y)^6 \; R_m^{6} \; A_H}
\label{c6}
\end{eqnarray}
Further as discussed in \cite{bb3}, 
\begin{eqnarray}
g_0 =  {\Omega_b \over \Omega_{\rm eff}} \approx {1 \over (1+y)^3} \; \left({R_0 \over R_m}\right)^3 =  {64 \; y^3 \over (1+y)^3}  
\label{g0}
\end{eqnarray}
Based on the stability analysis of amorphous systems structure, $z$ is predicted to be of the order of $3$ (for a three dimensional block).  Further, the intermolecular interactions being rather weak, they can mix very few single molecule levels. The dipole nature of these interactions further suggest  ${\mathcal N}=3$ (the  number of relevant vibrational energy levels in a molecule); this in turn implies $\eta=2$.
The above on substitution in eq.(\ref{vnpsa}) leads to 
\begin{eqnarray}
b  \approx   \; { 36  \over   \eta  \; \sqrt{ z \; g_0} \; A_H}  \; {y^6 \over (1+y)^6}
=  { 9  \over 4 \; \sqrt{ 3 } \; A_H }  \; \left({y \over 1+y}\right)^{9/2}
\label{b1}
\end{eqnarray}  
As clear from the above, $b$ is quite sensitive to the ratio $y= {R_v \over R_m}$, with $ R_v$ as half of the  distance  between two  nearest SRO molecules and $R_m$ as their radius. Experimental data of some standard glasses suggest that $R_v$ is typically of the same order as $R_m$  \cite{bb3}.  Hereafter  we use $y = {R_v \over R_m} \sim 1$ in our quantitative analysis. Eq.(\ref{g0}) then gives  $g_0 =8$.

 {\it Determination of $A_H$}: for materials in which spectral optical properties are not available, two refractive-index based approximation for $A_H$ namely, standard Tabor-Winterton  approximation (TWA) and single oscillator approximation (SOA),  provide useful estimates \cite{fr2000} . As indicated by previous studies, TWA is more appropriate for low refractive index materials (for $n < 1.8$);  $A_H$ in this case is given as (see eq.(11.14) of \cite{isra, fr2000})
\begin{eqnarray}
A_H \approx {3 \hbar \nu_e \over 16 \sqrt{2}} {(n_0^2-1)^2 \over (n_0^2 +1)^{3/2}} 
\label{ah}
\end{eqnarray}  
with $n_0$ as the refractive index at zero frequency and $\nu_e$ as the characteristic absorption frequency in the ultra-violet (also referred as the plasma frequency of the free electron gas, with a typical value is $\nu_e=3 \times 10^{15}$ for most ceramics). Further $n_0$ and $\nu_e$ in eq.(\ref{ah}) can be obtained by the standard routes (e.g. Cauchy's Plots or other available formulas) \cite{fr2000}: $n^2(\nu) -1  = {\mathcal G}_{uv} + {\nu^2 \over \nu_e^2} \; (n^2(\nu) -1)$. For cases, where the frequency-dependence of $n$ is known either as an exact formula (e.g. V52, BALNA, LAT) or available for two or more frequencies (see section VII of the supplemental part \cite{sup}), we determine $\nu_e$ and  and $n_0=\sqrt{1+{\mathcal G}_{uv} }$ by least square fit to plot $(n^2-1)$ vs $(n^2-1)/\lambda^2$. (Note eq.(\ref{ah}) is applicable for the case for two molecules interacting by VW interaction with vaccum as the intervening medium, with zero frequency contribution neglected; see eq.(11.14) of \cite{fr2000, isra}).

 For materials with higher indexes (for $ n > 1.8$) however  TWA  is found to be increasingly poor; instead, the single oscillator approximation (SOA) \cite{fr2000}  is closer to the exact values
\begin{eqnarray}
A_H \approx 312 \; \times  10^{-21} \; {(n^2-1)^{3/2} \over (n^2 +1)^{3/2}} \: J
\label{ahs}
\end{eqnarray}  
with $n$ as the available value of refractive index; for our calculation, we choose $n=n_0$.

Table I displays $n_0$ values along with $\nu_e$ values for the 18  glasses;  (although the latter is not used for the higher index cases but is included here  for the sake of completeness). The corresponding $A_H$ values obtained by either eq.(\ref{ah}) or eq.(\ref{ahs})  are also given in the table along with $b$ values from eq.(\ref{b1}).

\section{Density of States (DOS)}

  At very low temperatures, the induced dipole  interactions  result in excitations among vibrational energy levels of molecules (not strong enough to  excite the electronic states and the chemical bonding prevents  the rotation of molecules). In this section, we derive the ensemble averaged density of the states which participate in these excitations. The analytical tools however depend on the type. of the ensemble. As discussed in previous section,  the only information available about the ensemble  is that it corresponds to independent but randomly distributed matrix elements of ${\mathcal H}$ with first two moments given by eq.(\ref{hav1}),  eq.(\ref{vnp}). Here we consider two different approaches: the first one, a familiar route  based on the moments of the distribution, derives bulk DOS only but the second one, although technically complicated, is exact and provides information about the edge DOS as well as the higher order spectral correlations; (the latter are used in next section for the averaging of the partition function). The details of both approaches are discussed in supplementary part \cite{sup}).

 {\it (i) Arbitrarily distributed matrix elements:}    The many body DOS $\rho(e) =  \sum_n \delta(e-e_n) $  of  the eigenvalues $e_k$, $k=1 \to N$ of the Hamiltonian ${\mathcal H}$  (eq.(\ref{hb1}))  can be expressed in terms of the standard Green's function formulation: 
$\rho(e) = -{1 \over \pi} \; {\rm Im} \; G^{\dagger}(e)$ with $G^{\dagger}(e)=\lim_{\epsilon \to 0} G(e+i \epsilon)= \lim_{\epsilon \to 0}{\rm Tr} 
{1 \over {\mathcal H}-e - i \epsilon}$. 
The ensemble averaged density of states can then be written as  \cite{hjs}
\begin{eqnarray}
\langle \rho(e) \rangle=- {1 \over \pi} \;\lim_{\epsilon \to 0}  \; {\rm Im} \; \langle  \; G(z) \; \rangle
\label{rhoe}
\end{eqnarray}
with $z$ as a complex number: $z=e+i \epsilon$. 
One can further write $\langle G(z) \rangle$ in terms of the moments $T_n \equiv \langle {\rm Tr}\; {\mathcal H}^n \rangle $:
\begin{eqnarray}
\langle G(z) \rangle &=& \langle {\rm Tr} {1 \over {\mathcal H}-z} \rangle 
=-{1 \over z} \sum_{n=0}^{\infty} \; {1 \over z^n} \;  T_n
\label{g4}
\end{eqnarray} 
It is easy to calculate the first three moments. As discussed in section II, $\langle {\mathcal H}_{\mathcal K L} \rangle=0$, $\langle {\mathcal H}_{\mathcal KL}^2 \rangle = v_{\mathcal KL}$  and $\langle {\mathcal H}_{\mathcal KL} \; {\mathcal H}_{\mathcal MN}\rangle =0$   if  $({\mathcal K, L}) \not=({\mathcal M, N}) $ with $v_{\mathcal KL}$ given by eq.(\ref{vnp}).
This gives   $T_0= N$, $T_1=0$ and 
\begin{eqnarray}
T_2 =\sum_{\mathcal K, L}  v_{_{\mathcal KL} } &=&    {N\over  2 \; b^2} 
\label{t2}
\end{eqnarray}
where  $b$ is given by eq.(\ref{b1}). To obtain higher order traces, we expand $ {\rm Tr}\; {\mathcal H}^m $ in terms of the matrix elements,
\begin{eqnarray}
T_m = \langle  {\rm Tr}\; {\mathcal H}^m \rangle &=& \sum_{ {\mathcal K}_1, {\mathcal K}_2,.. {\mathcal K}_n}  \langle {\mathcal H}_{ {\mathcal K}_1 {\mathcal K}_2} \; {\mathcal H}_{ {\mathcal K}_2 {\mathcal K}_3} \; .....{\mathcal H}_{ {\mathcal K}_{m-1} {\mathcal K}_m} {\mathcal H}_{ {\mathcal K}_m {\mathcal K}_1} \rangle
 \end{eqnarray} 
 
As clear from the above, the trace operation ensures that the terms always have a cyclic appearance. 
 Further in evaluating $T_m$ for $N\rightarrow \infty$,  the terms with only pairwise (binary) correlations will be of consequence.  For materials where  ${\mathcal H}_{\mathcal K L}$ behaves as a Gaussian variable, this follows from the Wick's probability theorem or Isserlis theorem \cite{issr} (see supplementary part \cite{sup}). For ${\mathcal H}_{\mathcal K L}$ as a non-Gaussian variable, the above follows due to total number of terms with products of pairs being much larger than all other terms. For example, consider the contribution to $T_{2n}$ from the terms consisting of $n$-products of  $v_{\mathcal KL}$ (i.e of type $v_{\mathcal KL_1}\; v_{\mathcal L_1L_2} \ldots v_{\mathcal L_{n-1}K}$) with  ${\mathcal K, L}$ pairs as $2$-plets.  
From eq.(\ref{vnp}) and eq.(\ref{c6}), $v_{\mathcal KL} \sim A_H^2 \sim 10^{-38} \; J$. The contribution to $T_{2n}$ from  all terms with $n$ pairwise correlations among $2$-plets is then of the order of $N \times \left( {1\over 2} \;\eta^{2} \; g_0(g_0-1) \; 10^{-38}\right)^n \; J \approx  N \times 2^{7n} \; 10^{-38 n} \; J$ (with $g_0=8$ and $\eta=2$). Similarly the ${\mathcal K, L}$ pairs forming  higher order plets also contribute,  their weaker individual contribution compensated by higher number of terms. The total  
number of terms contributing to $T_{2n}$ and  forming $n$-products of  $v_{\mathcal KL}$ are $N \times \sum_{p_1,\ldots, p_n=2}^{g_0} {}^{g_0} C_{p_k} \; \eta^{p_k}$.
In contrast, the number of terms in $T_{2n}$ with one or more matrix elements appearing repeatedly is relatively much less (e.g. the terms of type $\langle \left( {\mathcal H}_{\mathcal KK}  \right)^{2n}\rangle$ are only $N$), their net contribution is different by an order of magnitude and can be neglected. (Note, with typical molecular vibrational energies $~10^{-20} \; J$ and $g_0 \approx 8$, the upper bound on ${\mathcal H}_{\mathcal KK}$ is $10^{-19} \; J$. This in turn gives $\sum_{\mathcal K} \; \left( {\mathcal H}_{\mathcal KK}  \right)^{2n} \sim N \times 10^{-38 n} \; J$). 
As discussed in section III of the supplementary part \cite{sup} in more detail, $T_{2n}$ can then be approximated as 
\begin{eqnarray}
T_{2n} &\approx & {1\over n+1} 
\left( {\begin{array}{cc}   2 n  \\  n \end{array} } \right)
\; {N \over  2^n \; b^{2 n}}
\label{t2n}
\end{eqnarray}

The terms contributing to $T_{2n+1}$  are of type  $(v_{\mathcal KL})^{2(n-t)} \; \langle ({\mathcal H}_{\mathcal KK})^{2t+1} \rangle$; their contribution can however be set to zero by an appropriate choice of the origin of single molecular energy scale. 
 For technical simplification and again without loss of generality, here we set $\varepsilon_0=0$ which, from eq.(\ref{mean}), results in $\langle {\mathcal H}_{\mathcal KK} \rangle=0$. Following same steps as above (see supplementary part \cite{sup}), this again leads to $\langle {\mathcal H}_{\mathcal KK})^{2t+1} \rangle =0$ and as a consequence \cite{sup}
 \begin{eqnarray}
T_{2n+1} \rightarrow  0
\label{m3}
\end{eqnarray}
 
Substituting the above results in eq.(\ref{g4}), we get 

\begin{eqnarray}
\langle G(z) \rangle &\approx & - {N \over z} \; \sum_{n=0}^{\infty} \; 
{1\over n+1} 
\left( {\begin{array}{cc}   2 n  \\  n \end{array} } \right)
 \; \left({1 \over  2 \; b^2 \; z^2}\right)^n    \nonumber \\
&=& - \; {N z b^2} \; \left[1- \left(1-{2  \over b^2 z^2} \right)^{1/2}\right]
 \label{g5}
\end{eqnarray}
Now substituting $z=e+i \epsilon$, taking the imaginary part of  the above, followed by limit $\epsilon \to 0$  then  gives 
\begin{eqnarray}
\langle \rho_{bulk}(e) \rangle  
&=&    {N  b \over  2\pi } \;  \sqrt{2 - \left({b e}\right)^2}.
\label{rhoe01}
\end{eqnarray}
As clear from the above, the bulk of the spectrum, with its width as $2\sqrt{2}/b$ and mean level spacing $\Delta_b \approx {\pi \sqrt{2} \over N b}$,  depends on the single parameter i.e the {\it bulk-spectrum parameter} $b$ given by eq.(\ref{b1}).  As $b$ depends on the  average properties of the many body inter-molecular interactions, it is not expected to vary much from one system to another. This is also indicated by eq.(\ref{b1}). 
For later use, we shift the origin of the spectrum to $e=-\sqrt{2}/b$  and redefine $b=b_0 \sqrt{2}$ which gives 
\begin{eqnarray}
\langle \rho_{bulk}(e) \rangle  
&=&    {N  b_0 \over  \pi } \;  \sqrt{b_0 e (2-b_0 e)} 
\label{rhoe1}
\end{eqnarray}
As displayed in table I for 18 non-metallic glasses, $b \sim 10^{17}-10^{18} \; J^{-1}$.

Although the derivation given above does not assume any specific distribution of the matrix elements of the Hamiltonian, eq.(\ref{rhoe1}) is analogous to the bulk level-density of a Gaussian orthogonal ensemble (GOE). The latter refers to an  ensemble of real-symmetric matrices with independent Gaussian distributed matrix elements with zero mean and  variance of the diagonal twice that of the off-diagonal.  This is not surprising, a semi-circle behavior of the bulk level density is known to be valid  for a wide range of complex systems described by the matrix elements distributions with finite moments, (typical of dense matrices, irrespective of the nature of their randomness) (see section 4.3 of \cite{me} and \cite{thou,bray}). As in our case,  the total number of off-diagonals  ($\approx  N \; \sum_{p=1}^{g_0}  \; ^{g_0} C_p \; \eta^p$) is much larger than the diagonals (total $N$ of them), $H$ {\it effectively} behaves as a dense matrix and a semi-circle behavior of the level density is expected.

{\it (ii) Gaussian distributed matrix elements:} 
The definition of $\rho(e)=\sum_n \delta(e-e_n)$ gives  $\langle \rho(e) \rangle =N \int P_e \; {\rm d}e_2 \ldots {\rm d}e_N $ with  $P_e \equiv P_e(e_1, e_2,\ldots, e_N)$ as the joint probability density function (JPDF) of the eigenvalues $e_k$, $k=1 \to N$. This leads to an alternative route to derive  $\langle \rho(e) \rangle$: it  is based
on first deriving a diffusion equation for $P_e$ and thereby $\langle \rho(e) \rangle$ with changing ensemble parameters 
followed by its solution for their desired values. Although applicable for non-Gaussian ensembles too \cite{psco}, here we describe it only for then cases when ${\mathcal H}$ described by a multiparametric Gaussian ensemble \cite{psall}. The steps are as follows.

As mentioned in section II, both $C^{(3n)}_{\mathcal KL}$ and ${\mathcal A}^{(a,b)}_{\mathcal KL}$ behave as random variables, with finite mean and variance, over an ensemble of the basic block replicas.  In absence of any information about their higher order moments,  one can invoke  standard maximum entropy hypothesis and assume both  to be  Gaussian variables. Further as (i) each $E_{{\mathcal K}_n}$, the energy of a SRO molecule, by itself is  a sum over many atomic energy levels with finite energies, (ii) each atom consists of many particle interactions,  it is appropriate to assume a randomization of the atomic energy levels and thereby $E_{{\mathcal K}_n}$ (see {\it appendix A} and also \cite{gmw,zkpcd}).  Following standard central limit theorem, each $E_{\mathcal K}=\sum_{n=1}^{g_0} E_{{\mathcal K}_n}$ can then be assumed  to be Gaussian distributed over an ensemble of the replicas of SRO molecules (even though $g_0 \sim 10$). This in turn permits ${\mathcal H}$ to be modeled by a multiparametric Gaussian ensemble of real-symmetric matrices with independent elements, described by the ensemble density 
\begin{eqnarray}
\rho_{{\mathcal H}} ({\mathcal H})=C \; {\rm exp}[{- \sum_{k\le l} {1 \over 2 \nu_{kl}} ({\mathcal H}_{kl}-b_{kl})^2 }]
\label{rhog}
\end{eqnarray}
Here the mean $b_{kl}=\langle {\mathcal H}_{kl} \rangle$ and variance $\nu_{kl}=\langle {\mathcal H}_{kl}^2 \rangle-\langle {\mathcal H}_{kl} \rangle^2$  are given by  eq.(\ref{hav1}) and eq.(\ref{vnp}). 

As reviewed briefly in section IV of the supplementary part \cite{sup} and discussed in detail in \cite{psall}, an arbitrary variation of the parameters $\nu_{kl}, b_{kl}$ subjects $\rho_{{\mathcal H}}$ and thereby $P_e$ and $\langle \rho(e) \rangle$ to evolve  from an arbitrary initial ensemble; the evolution however is  governed by  a single parameter $Y$, referred as the complexity parameter \cite{psall}:
\begin{eqnarray}
Y= -{2\over  \gamma \; N(N+3)}  \; \; {\rm ln}\left[ \prod_{k \le l} \;|1 - (2- \delta_{kl})  \gamma \; v_{kl}| \quad |b_{kl} + b_0|^2 \right] + constant
\label{y}
\end{eqnarray}
with $\gamma$ as an arbitrary parameter (marking the end of the evolution) and $b_0=1,0$ if $b_{kl}=0$ or $\not=0$ respectively.  Here the initial state  of evolution is labelled by $Y=Y_0$.  In $\lim_ {Y \to \infty}$, both $\rho_H$ as well as $P_e$ and thereby $\langle \rho(e) \rangle$ approach the GOE limit (see supplementary part \cite{sup} for more details  or \cite{psall}.

To derive $\langle \rho(e)\rangle$ for the basic block ensemble (\ref{rhog}), we first need to determine its complexity parameter $Y$. A substitution of  eq.(\ref{mean}) and eq.(\ref{vnp}) in eq.(\ref{y}) gives 
\begin{eqnarray}
Y = {-2\over \gamma N(N+3)} \left[\sum_{k,l \atop k< l}\ln |1- 2 \; \gamma \; v_{kl}| + \sum_k \ln|1-\gamma \; g_0 \; \nu_0| + \sum_k \ln|g_0 \; \epsilon_0| \right].
\label{yg}
\end{eqnarray}
 The initial state  here corresponds to  the ensemble of diagonal ${\mathcal H}_0$ matrices (given by eq.(\ref{h1kl})) with $Y_0= {-2\over \gamma N(N+3)} \left[ \sum_k \ln|1-\gamma \; g_0 \; \nu_0| + \sum_k \ln|g_0 \; \epsilon_0 | \right]$. This in turn gives 
\begin{eqnarray} 
Y-Y_0={-2\over \gamma N(N+3)} \; \sum_{k,l \atop k< l}\ln |1- 2 \;\gamma \; v_{kl}| \approx {4 \over N(N+3)} \sum_{k,l \atop k< l} v_{kl}
\label{yg1}
\end{eqnarray}
Using eq.(\ref{vnpsa})  in the above gives $Y-Y_0 \approx {2 \over N b^2}$. With typically $b \sim 10^{18} \ J^{-1}$ (see table I) and $g_0 \nu_0 \sim 10^{-67} \; J^2$ (see section II), this implies  $Y \gg  Y_0$ and, as discussed in supplementary part \cite{sup} and \cite{psall}, the evolution of $P_e$ and thereby  $\langle \rho_e(e)\rangle$ approaches  the stationary limit of a  GOE throughout the spectrum i.e both bulk as well edge. Note however same is  not true for the higher order spectral correlations as they evolve with different speed (see supplementary part \cite{sup}).

As discussed in detail in \cite{psco}, a complexity parametric formulation of  $P(e_1, e_2,\ldots, e_N)$ can  be derived for non-Gaussian cases too; as expected, $Y$ for these case is more complicated. However here again $\langle \rho(e) \rangle$ approaches GOE limit  as $Y \gg Y_0$. (Note here the initial ensemble and the evolution equation for the $\langle \rho(e)\rangle$ remains  same as in the previous case).

Although the first route discussed above i.e the determination of $\langle \rho(e) \rangle$ from its moments, gives the behavior only  in the bulk, the second route, based on the complexity parameter formulation, gives the behaviour for all energy ranges. 
Following the above reasoning, the ensemble averaged  edge level density $\langle \rho_{edge}(e) \rangle$ for ${\mathcal H}$ can then be modeled by that of a GOE too. Using a generic form, one can write
\begin{eqnarray}
\langle \rho_{edge-t}(e) \rangle = {N \;\; b_0 \over \sqrt{\lambda}} \; f_t(\lambda \; b_0 \; e). 
\label{rlt}
\end{eqnarray}
 with subscript $t = L, U$ denoting lower ($- \infty  < x <0$) and upper edge ($2 <  x < \infty$), respectively, and $f_t(x)$ in case of a GOE is
 
\begin{eqnarray}
f_L(x) & \approx & x \; Ai^2(-x)  + (Ai'(-x))^2 + {1 \over 2} \;  Ai(-x) \; E(-x)     \hspace{0.3in} ({\rm at \; lower \;  edge})   \label{a1l} \\
f_U(x) &\approx&   - (x-2) \; Ai^2(x-2)  + (Ai'(x-2))^2  + {1\over 2} \;  Ai(x-2) \; E(x-2)  \hspace{0.1in} 
({\rm at \; upper \;  edge})
\nonumber \\
\label{a1u}
\end{eqnarray}
with $E(x)  \equiv  \int_{-\infty}^{x} Ai(y) \; {\rm d}y $,  with $Ai(y)$ as the Airy function of the first kind \cite{airy}.  For later use, it is worth noting that Airy-function asymptotic of eqs.(\ref{a1l},\ref{a1u}) leads to a super-exponential form for $f_L(x)$ for $x < 0$ ({\it appendix B})
\begin{eqnarray}
f_L(x) \sim {1 \over \sqrt{|x|}} \; {\rm e}^{-\sqrt{|x|^3}}.
\label{fx}
\end{eqnarray}
and a square-root form for $x \gg 0$,
\begin{eqnarray}
f_L(x) \approx {\sqrt{c_0^2+x}\over \pi} + {1\over \pi x} \; {\rm cos}\left({2 \over 3}\; x^{3/2}\right).
\label{rapp1}
\end{eqnarray}
where $c_0=0.1865 \; \pi$.
As expected $f(x)$ is non-zero at $x=0$; (note $f(x)=0$ would imply a gap at $x=0$ instead of smooth connection with the bulk density). Substitution of eq.(\ref{rapp1}) in eq.(\ref{rlt}) gives 
\begin{eqnarray}
\langle \rho_{edge-l}(e) \rangle \approx {N \; b_0 \over \pi \sqrt{\lambda}} \; \sqrt{c_0^2 + \lambda b_0 e}  
\hspace{0.1in} (e >0)
\label{rhop}
\end{eqnarray} 
Thus, for $e >0$, the edge behavior  smoothly connects with the beginning of bulk of the ensemble averaged level density (see eq.(\ref{rhoe1}) and figure 1).

Contrary to bulk, the edge-density depends on two parameters $b$ and $\lambda$; the latter, governing the decay of density of states in the lower edge, can be referred as the {\it edge-spectrum parameter}. 
Here $\lambda$ is  dimensionless and satisfy the requirement that $\langle \rho_{edge}(e) \rangle= \langle \rho_{bulk}(e) \rangle$ near $e \sim e_0$. 
For the GOE case, we have
\begin{eqnarray}
\lambda= \lambda_0 \; N^{2/3} =   {\mathcal N}^{2 g_0/3}  \; \lambda_0 = 3^{16/3} \;  \lambda_0 \;
\label{lamb}
\end{eqnarray}
with $\lambda_0$ is a constant: $\lambda_0 \approx \sqrt{2}$.  The $2^{nd}$ equality in the above follows from  ${\mathcal N}=3$ and $g_0 \approx 8$ (see text below eq.(\ref{g0}).
For later use, it is necessary to determine the point, say $e_0$, where the edge of the spectrum meets the bulk; as discussed in {\it appendix B} and also clear from figure 1, the normalization condition $\int_{-\infty}^{\infty} \langle \rho_e \rangle \; {\rm d}e = N$ gives 
\begin{eqnarray}
e_0 \approx  {1\over   3 \; b_0 \lambda } \sim  10^{-21} \; J
\label{e0t}
\end{eqnarray}

Following from eq.(\ref{fx}), the rapid decay of density of states  for $ e < 0$ implies very few levels in the edge region; this can directly be  confirmed from eq.(\ref{a1l}) which gives their number, say $N_{edge}$ as 
\begin{eqnarray}
N_{edge}(e)= \int_{-\infty}^e  \langle \rho_{edge-l}(e) \rangle \; {\rm d}e  = {N \;{\mathcal F}(\lambda b_0 e)  \over \sqrt{\lambda^3} }={{\mathcal F}(\lambda b_0 e) \over \sqrt{\lambda_0^3}} . 
\label{nlt}
\end{eqnarray}
with ${\mathcal F}(x)=\int_{-\infty}^{x} \; {\rm d x} \;  f_L(x)$. As can be seen from figure 2, ${\mathcal F}$ is maximum at $e=0$,  with ${\mathcal F}(0) \approx 0.16$, and rapidly decays as $|e|$ increases. Clearly most of the contribution to $f_0$ is coming from the region $x \sim 0$, indicating the presence of almost all levels in the lower edge region very close to $e=0$. Further the mean level spacing $\Delta(e)$ at an arbitrary energy $e$ in the edge region is 
\begin{eqnarray}
\Delta(e) = {1\over \langle \rho_{edge-l} \rangle} \approx  {\sqrt{\lambda} \over N \;  b_0 \; f_L(\lambda b_0 e)}.
\label{delb} 
\end{eqnarray}
The above indicates $\Delta(0) \ll \Delta(e_l)$ with $e_l \ll 0$.

\section {Heat Capacity of the basic block}

The heat capacity of a thermodynamic system at a temperature $T$ can be described in terms of its partition function $Z = \sum_{n=1}^N  {\rm exp}[-\beta e_n] $ with $e_n$ as its energy levels and $\beta = 1/(k_b T)$ with $k_b$ as the Boltzmann constant. But 
the energy levels i.e the eigenvalues $e_n$ of the Hamiltonian of a complex system fluctuate, in general, from one sample to the other and therefore $Z$ is also subjected to the sample to sample fluctuations. As a consequence, it is a necessary as well as a standard procedure to consider the averaging of $Z$ over exact replicas of the system (referred as ensemble or disordered averaging).

An amorphous system with its many body interactions is a complex system.
Our next step is to calculate the ensemble averaged heat capacity  $C_v$ of a basic-block of volume $\Omega_b$, defined as $\langle C_v \rangle  =  k \; \beta^2 \; {\partial^2 \over \partial \beta^2} \; \langle  {\rm log } \; Z \rangle$ with $Z$ as its partition function: $Z = \sum_{n=1}^N  {\rm exp}[-\beta e_n] $. For cases with annealed disorder, 
\begin{eqnarray}
\langle  {\rm log } \; Z \rangle = {\rm log } \langle Z \rangle.
\label{loz}
\end{eqnarray}
 Although disorder in glasses is generally believed to be of quenched type, the belief  is based on the experimentally observed long range structural disorder at macroscopic scales. No such evidence is available however in case of nanosize samples. Further as there is ample evidence of thermodynamics in nanoscales being different from macroscales (see e.g. \cite{nano1} and references therein), a quenched behavior in the latter  does not imply the same in the former. The nature of disorder is also different in  the two cases. In contrast to a fixed structural disorder at macroscopic samples, the disorder in the basic block  arises due to rapidly changing orientation of the instantaneous dipoles and can appropriately be considered as annealed type (with no external impurities but complexity of interactions leading to randomization of dynamics). 

Alternatively, following replica trick, one can write $\langle  {\rm log } \; Z \rangle=\lim_{n \to 0} \; {1\over n} \log \langle Z^n \rangle$. But, as discussed in section V of the supplementary part \cite{sup},  the local correlations of the  ${\mathcal H}$ ensemble  in the spectral edge are almost negligible; this in turn leads to $\langle Z^n \rangle \approx  \langle Z \rangle^n$ and thereby the relation (\ref{loz}). Intuitively this can be explained as follows: at very low temperature, the Hamiltonian ${\mathcal H}={\mathcal H}_0+{\mathcal U}$ is in energy states in the lower spectrum edge. Although the perturbation ${\mathcal U}$ affects the average density of states in the region but is not strong enough to alter the local density correlations which  remain almost same as their  initial state ${\mathcal H}_0$. As the latter in our case corresponds to  a sum over free molecule Hamiltonians with independent energies, the  energy levels of ${\mathcal H}$ near the edge are uncorrelated which in turn ensures the relation (\ref{loz}). This  in turn gives 
\begin{eqnarray}
\langle C_v \rangle  &=&  k \; \beta^2 \; {\partial^2 \over \partial \beta^2} \; {\rm log } \langle  Z \; \rangle
\label{cv1}
\end{eqnarray}

With $\rho(e)$ as the density of states $\rho_e(e)=\sum_{n=1}^N  \delta(e-e_n)$ of a typical basic block, $Z$ can also be expressed as $ Z =  \int {\rm d e} \;  \rho_e(e)  \;   {\rm exp}[-\beta e]$. 
For energy ranges where level-spacing is very small, $\rho(e)$ can be approximated by a smooth function i.e its spectral average. In  the  present case, however, the spectrum consists of regions with very large mean level spacing for $e <0$ and  it is appropriate to separate the discrete and continuous parts of $Z$ \cite{dive}: $Z =    \sum_{n; e_n < 0}  {\rm exp}[-\beta e_n] +  \int_0^{\infty} {\rm d e} \;  \overline{ \rho_e}\; {\rm exp}[-\beta e]$ with $\overline{ \rho_e}$ as the spectral averaged density of states. This in turn leads to
\begin{eqnarray}
 \langle  Z  \rangle  =   \sum_{n; e_n <0}  \langle {\rm exp}[-\beta  e_n ] \rangle +  \int \; {\rm d e} \;  \langle \overline{\rho_e} \rangle    \; {\rm exp}[-\beta e]
\label{zz}
\end{eqnarray}
with  $\langle \overline{\rho_e}\rangle$ as the  level density averaged over the spectrum as well as the ensemble. (Note the first term here is not spectral averaged).

\subsection { Calculation of $\langle Z \rangle$}

Due to different  functional behavior of the level density in the edge and bulk regions, we  divide $\langle Z \rangle$ in four  parts corresponding to lower edge ($ - \infty < e \le e_0$) ,  bulk ($e_0 \le e \le (2/b)-e_0$)  and upper edge ($(2/b)-e_0 < e \le \infty$)  respectively:
\begin{eqnarray}
\langle Z \rangle 
&=&   \sum_{n; e_n <0}  \langle {\rm exp}[-\beta  e_n ] \rangle + J_L  + J_B + J_U 
\label{z1}
\end{eqnarray}
where 
\begin{eqnarray}
J_L   & =&  \int_{0}^{e_0} \; {\rm d e} \; \langle \rho_{edge-l}(e) \rangle   \;  \; {\rm exp}(-\beta \; e) 
\label{jl0}   \\
J_B   & =&    \int_{e_0}^{{2\over b_0}-e_0} \; {\rm d e} \;  \langle \rho_{bulk}(e) \rangle    \;  \; {\rm exp}(-\beta \; e)
\label{jb0}  \\
J_U   & =&    \int_{{2\over b_0}-e_0}^{\infty} \; {\rm d e} \;  \langle \rho_{edge-u}(e) \rangle   \;  \; {\rm exp}(-\beta \; e) 
\label{ju0}
\end{eqnarray}

Substituting eq.(\ref{rhop}) in eq.(\ref{jl0}), one can rewrite it  as 
\begin{eqnarray}
J_L  &=& {N \; b_0\over \pi \sqrt{\lambda}}  \int_{0}^{e_0} \; {\rm d e} \;  \sqrt{c_0^2 + \lambda b_0 e}  
 \; \;  {\rm exp}(-{\beta \; e } ) \label{jl1} \\
  &\approx & {N \over \pi} \; \left({1 \over 3 \lambda \beta e_0}\right)^{3/2}  \;  
\left[ \gamma\left({3\over 2},  \left(1+ \eta \right) \beta e_0 \right) -
\gamma\left({3\over 2}, \;\; {\eta \beta e_0}  \right)  \right] \; {\rm exp}[\eta \beta e_0]
\label{jl}
\end{eqnarray}
where $\eta \approx 1$,  $c_0^2 =  (0.1865 \pi)^2 \approx 1/3$ and  and  $\gamma(a, x)$ is the incomplete Gamma function defined as $\gamma(a,x) = \int_0^x \; t^{a-1} \; {\rm e}^{-t} \; {\rm d}t$ with $e_0$ and $\lambda$ given by eq.(\ref{lamb}), eq.(\ref{e0t}) respectively.

Similarly,  $J_B$  can be written as (with $\langle \rho_{bulk} \rangle$ given by eq.(\ref{rhoe1}))
\begin{eqnarray}
J_B   
 &\approx &    {N \over \pi}  \; \int_{be_0}^{2-b e_0}  {\rm d x} \;  \sqrt{x (2-x)} \; {\rm e}^{-{\beta \over b}  x}  \label{jbt6-} \\
&=&  {N\over 2\pi} \;\;  \Phi \left({3\over 2}, 3, - {6 \lambda \beta e_0}\right) -  {N\sqrt{2}  \over \pi}  \;\; \left({1 \over 3 \lambda \beta e_0} \right)^{3/2} \; \gamma\left({3\over 2}, \beta e_0 \right)-{N\sqrt{6 \lambda-1}\over 9 \pi \lambda^2 \beta e_0 } \;  \; {\rm e}^{-{6 \lambda \beta e_0}} \nonumber \\
\label{jbt6} 
\end{eqnarray}
with $\Phi(a,b,x)$ as the confluent Hypergeometric function, defined as $\Phi \left(a,b_0;x\right) = \int_{0}^{2}  {\rm d x} \; \sqrt{x \; (2-x)} \;  {\rm e}^{-{\beta \over b_0}  x} $ (see section VI of the supplementary part \cite{sup}).
Due to presence of the term  ${\rm e}^{-{\beta \over b}  x}$  in the integrant, the contribution from the above integral is significant when $\beta e_0 <1$  (i.e  $k_B T > e_0$ or the temperature $T$ is high enough to ensure the thermal perturbation to access the states in the bulk). 
With $e_0 \approx (3 b_0 \lambda)^{-1}$, this requires $T > (3 k_B b \lambda)^{-1} \sim 75^o \; K$. (with $\lambda \approx   495 $, $b \sim 10^{18} J^{-1}$ and $k_b = 1.38 \times 10^{-23} \; J/K$).

Further  $J_U$ can  be calculated  by substituting eq.(\ref{rlt}) with $f_U$ given by eq.(\ref{a1u}). As the  integration-range   now is $ 2-b e_0 < x < \infty$, the level-density decreases faster than exponential in this range and one can write  
\begin{eqnarray}
J_U \approx {\rm e}^{-3 \beta e_0} \; J_L. 
\label{ju}
\end{eqnarray}
This leaves the contribution from $J_U$  significant only for large $3 \beta e_0 < 1$ or $T > o(10^2) \; K$.

Substitution of eqs.(\ref{jl},\ref{jbt6},\ref{ju}) in eq.(\ref{z1}) gives the partition function for the basic block
\begin{eqnarray}
\langle  Z \rangle &=& \sum_{n; e_n <0}  \langle {\rm exp}[-\beta  e_n ] \rangle   +  {N \over \pi} \; \left({1 \over 3\lambda  \beta e_0}\right)^{3/2}  \;  \left[ \gamma\left({3\over 2},  \left(1+ \eta \right){\beta e_0}\right) -\gamma\left({3\over 2}, \; {\eta \beta e_0}  \right)\right] \; {\rm e}^{\eta \beta e_0}  + .\nonumber \\&+&  {N\over 2\pi} \;\;  \Phi \left({3\over 2}, 3, - {6 \lambda \beta e_0}\right) -  {N\sqrt{2}  \over \pi}  \;\; \left({1 \over 3 \lambda \beta e_0} \right)^{3/2} \; \gamma\left({3\over 2}, \beta e_0 \right)-{N\sqrt{6 \lambda-1}\over 9 \pi \lambda^2 \beta e_0 } \;  \; {\rm e}^{-{6 \lambda \beta e_0}}
\label{jbj7} 
\end{eqnarray}

A substitution of the above expression in eq.(\ref{cv1}) gives, in principle, the specific heat. The above expression can further be simplified by noting that 
(i) only a single level is present in the tail region $e <0$ and that too is very close to $e \sim 0$, one can approximate $\sum_{n; e_n <0}  \langle {\rm exp}[-\beta  e_n ] \rangle \approx  1$,  
(ii) for $x \gg 1$, $\Phi \left({3\over 2}, 3, - x\right)  \approx  {2 \over \sqrt{\pi}} {1\over \sqrt{x^3}}$ and $\gamma\left({3\over 2},   x \right) \approx  \Gamma\left({3\over 2}\right) -\left(x\right)^{1/2} \; {\rm e}^{-x}$. 
As, in the present work, our interest is in temperature regime $T < 100^o {\rm K}$, this implies $\lambda \beta e_0 \gg  10^2$ (with $e_0$ given by eq.(\ref{e0t})) and  one can then  approximate 
\begin{eqnarray}
\langle  Z \rangle & \approx &  1  +  {N \over \pi} \; \left({1 \over 3 \lambda \beta e_0}\right)^{3/2}  \;  \left[ \gamma\left({3\over 2},  \left(1+\eta \right){\beta e_0}\right) -\gamma\left({3\over 2}, \; {\eta \beta e_0}  \right)\right] \; {\rm e}^{\eta \beta e_0}  
\label{jbj6} 
\end{eqnarray}

For comparison with the experiments however it is helpful  to analyze the specific heat in different temperature regimes.    

\subsection { $\langle C_v \rangle$ for low temperatures $T$}

\vspace{0.3in}

\noindent{\underline{\bf Case (a): ${\beta e_0} \gg 1$}}:
\vspace{0.3in}
As  $k_b T \ll e_0$ here,  the thermal perturbation mixes very few  states  even in the lower edge. With $\beta e_0  \gg 1$ in this case,  both $\gamma$-functions in eq.(\ref{jbj6}) can be expanded asymptotically as  
\begin{eqnarray}
\gamma\left(a,  x\right) \approx \Gamma\left(a\right)- x^{a-1} \; {\rm e}^{-x} \sum_{m=0}^{M-1} {(-1)^m \Gamma{(1-a+m)} \over \Gamma(1-a) \; x^m } + O(x^M)
\label{gm1}
\end{eqnarray}
Following from the above, the  2nd and higher order terms in the series for $\gamma\left({3\over 2},  \left(1+\eta \right){\beta e_0}\right)$ are smaller than those of $\gamma\left({3\over 2},  \eta {\beta e_0}\right)$ by an exponential factor and one can approximate  $\gamma\left({3\over 2},  \left(1+\eta \right){\beta e_0}\right) \approx \Gamma\left(3/2\right)$. This along with eq.(\ref{gm1}) reduces eq.(\ref{jbj6}) as
\begin{eqnarray}
\langle Z \rangle  \approx  1+  {\sqrt{\eta^3} \over \pi \; \sqrt{27 \; \lambda_0^3}  }    \;   \sum_{m=0}^{M-1} {(-1)^m \Gamma{(m-1/2)} \over \Gamma(-1/2) \; (\eta \beta e_0)^{m+1} } + O((\beta e_0)^M)
\label{zu0}\end{eqnarray}

Substitution of eq.(\ref{zu0}) in eq.(\ref{cv1}) now leads to, 

\begin{eqnarray}
C_{v}(T)  & \approx & {k_b \; \sqrt{\eta^3}  \over \pi \; \sqrt{27\;  \lambda_0^3}}   \; \sum_{m=0}^{M-1} {(-1)^m (m+1)(m+2) \Gamma{(m-1/2)} \over \Gamma(-1/2) \; (\eta \beta e_0)^{m+1} } + O((\beta e_0)^M)
\label{cv4u}
\end{eqnarray}
with $\eta \approx 1, \lambda_0 \approx \sqrt{2}, e_0  \sim 10^{-21} \; J$.  At very low temperatures, the above indicates a linear $T$-dependence of the leading order term: 
\begin{eqnarray}
C_{v}(T)  & \approx &   0.22 \; (\lambda \; b_0) \; k_b^2  \; T + 0.37  (\lambda \; b_0)^2 \; k_b^3  \; T^2 - 2.32 (\lambda \; b_0 )^3 \;k_b^4 \; T^3 + O(T^4)) 
\label{cvu}
\end{eqnarray}

It must be noted that the result above is based on approximating  $\gamma\left({3\over 2},  \left(1+\eta \right){\beta e_0}\right)$ only by its first term in the series expansion. At very low $T$ however the difference $\gamma\left({3\over 2},  \left(1+\eta \right){\beta e_0}\right)$-$\gamma\left({3\over 2},  \left(\eta\beta e_0\right)\right)$ rapidly decreases and the approximation is  not very good.  As discussed in  {\it appendix C}, a variation of $T$ leads to a transition in the behaviour of the leading order term of $C_v(T)$ from $T$ to $T^{3/2}$ which could well appear as an intermediate power in measurements.

\vspace{0.2in}

\noindent{\bf\underline{Case (b):  ${\beta e_0}  \ll 1$}}

\vspace{0.1in}

In limit ${\beta e_0 } \ll 1$, eq.(\ref{jl}) can be approximated by the behavior of $\gamma(a,x)$ near $x =0$: 
\begin{eqnarray}
\gamma(a,x) = \sum_{n=0}^{\infty} \; (-1)^n {x^{a+n} \over n! \; (a+n)}.
\label{gm2}
\end{eqnarray}
Substitution of the above in eq.(\ref{jbj6}) gives 
\begin{eqnarray}
\langle Z \rangle  
\approx   1+  {\sqrt{2} \; {\rm e}^{\eta \beta e_0} \over \pi \; \sqrt{27 \; \lambda_0^3}  }  \;  \sum_{m=0}^{M-1} a_m \; (\beta \; e_0)^m  + O((\beta e_0)^M) 
\label{zub}
\end{eqnarray}
with $a_m ={(-1)^m \over m! (m+3/2) } \; ((1+\eta)^{m+3/2}-\eta^{m+3/2}) $. 
Substitution of the above in eq.(\ref{cv1}) then leads to
\begin{eqnarray}
C_v   &\approx &  {k_b  \; {\rm e}^{\eta \beta e_0} \over \pi\; \sqrt{27 \; \lambda_0^3} \; }  
  \;  \sum_{m=0}^{M-1}  {b_m  (\beta e_0)^{m-2}} +O\left((\beta e_0)^{(M+1)}\right)
\label{cvdb}
\end{eqnarray}
with $b_m= {(-1)^m \over m!}\left[{\eta_0(m+2) \over (m+7/2)} - {2 \eta_0(m+1) \over (m+5/2)} + {\eta_0(m) \over (m+3/2)} \right]$ and $\eta_0(m) =2^{m+3/2}-1$.  The above gives the leading order term decaying as 
${1\over T^2}$.

\subsection{Comparison with experiments}

 Specific heat experiments on a wide range of glasses indicate a super-linear dependence on temperature below $T \leq 1^o K$: $c_v \sim T^{1+\varepsilon}$, $\varepsilon \sim 0.1 \to 0.3$ and a bump in the plot $c^{total}_v/T^3$  vs  $T$ in the region near $T \sim 10^o K$. Although the available experimental results are in general applicable for glass solids of macroscopic size, it is  tempting to compare them for those of microscopic size too.  Previous studies have indicated that the thermal properties of solids at nano scales are in general different from macro scales. More specifically, the specific heat at nano scales is expected to be bigger than that of macro scales \cite{nano1}.

  Theoretical results  mentioned in previous section are derived for the heal capacity of  a basic block of volume $\Omega_b$. The specific heat corresponding to  non-phononic contribution  can then be given as 
\begin{eqnarray}  
 c_v={1\over \rho_m \Omega_b} \; \langle C_v \rangle.
 \label{scv}
 \end{eqnarray} 
 with $\rho_m$ as the mass-density of the glass.
   As discussed in \cite{bb3,qc1},  $\Omega_b$ can be expressed in terms of the molar mass $M$ of the SRO molecule  participating in dispersion interaction: 
\begin{eqnarray}
\Omega_b = {64 M \over \rho_m \; N_a}
\label{omb}
\end{eqnarray}
 with $N_a$ as the Avogrado number.
 
 Experimental results on the macroscopic size glass material in general include the contribution from phonons too, with total $c_v$ given as
 \begin{eqnarray}  
 c_v^{total}= c_v + c_v^{ph}
 \label{scv1}
 \end{eqnarray} 
 where, for $T < T_D$,
\begin{eqnarray}
c_v^{ph} 
&=&  {9 R\over M} \;  \left({T \over T_D}\right)^3 \; \int_0^{x_D} {\rm d}x \; {x^4 \; {\rm e}^{x}\over ({\rm e}^x-1)^2}  \label{cvpht}\\
&\approx &   {R \over M \; x_D^3} \; \left[{12 \pi^4\over 5} - 9 (x_D^4 + 4 x_D^3 + 12 x_D^2 +24 x_D +24) 
{\rm e}^{-x_D}\right]  
\label{cvph}
\end{eqnarray}
 with $x_D=T_D/T$ and $T_D$ as the Debye temperature and $R=k_b N_a=8.31$ as the gas constant. Note here eq.(\ref{cvph} is obtained by rewriting the integral in eq.(\ref{cvpht}) as $I_1- I_2$ where $I_1=\int_0^{\infty} {\rm d}x \; f(x)$ and $I_2 =\int_{x_D}^{\infty} {\rm d}x \;f(x)$ with $f(x)= {x^4 \; {\rm e}^{x}\over ({\rm e}^x-1)^2}$. $I_1$ gives the standard low $T$ Debye result i.e $I_1= {12 \; \pi^4 \; R \over 5 \; M \; x_D^3}$. As for $I_2$, $x > x_D$, it can be approximated as $\int_{x_D}^{\infty} {\rm d}x \; x^4 {\rm e}^{-x}$ which in turn leads to 2nd term inside $( \; )$ in eq.(\ref{cvph}).

Although the dynamics of the phonons in a nanosize material is believed to be different, 
recent studies \cite{nano1}  suggest the validity of Debye formulation for $c_v^{ph}$ albeit with a different  $T_D$ value. This motivates us to consider the total $c_v$ for the basic block also described by eq.(\ref{scv1}) with phonon contribution given by eq.(\ref{cvph}). As discussed below, the experimental data for macrosizes  indeed agrees well  theoretical results for the nanosizes, especially at higher temperature, if eq.(\ref{scv1}) and eq.(\ref{cvph}) are used for the latter. More specifically $I_2$ needs to be retained for good agreement with experiments for $T > 10^o \; K$.
 
The heat capacity formulation derived in previous section are expressed in terms of the mathematical functions and their correspondence with experimental results is not directly obvious.  For numerical analysis and comparison with experiments, it is instructive to directly substitute eq.(\ref{jbj7}) in eq.(\ref{cv1}) and use computational techniques  to obtain  $c^{total}_v$ from eq.(\ref{scv1}) and eq.(\ref{cvph}).

\vspace{0.1in}

{\bf Comparison below $T \sim 1^o \; {\bf K}$}:
From eqs.(\ref{cvu},\ref{scv},\ref{omb}),  $c_v = c_1 \; T + o(T^2)$ with $c_1 \approx {0.22 \; \lambda \; b_0 \; k_b^2 N_a\over 64 M}$. The $c_1$ values for $18$ glasses are given in table I along with available experimental data (taken from \cite{phil}) for six glasses. While the agreement is good for $SiO_2, As_2S_3, Se$ and $PS$, the deviation in the case of  polymers PMMA and PC could be due to errors in correct estimation of (i)  mass $M$ of the SRO molecule, (ii) $y$ (ratio ${R_v\over R_m}$) in eq.(\ref{b1}), (ii) Hamaker constant $A_H$ (see eq.(\ref{ahs}). Note the first two reasons are more likely to affect the $c_v$ especially in case of polymers. Furthermore as the glass properties are quantitatively sensitive to the cooling process, a comparison with experiments requires a detailed knowledge of the glass-preparation history. (The $c_v$  in amorphous materials  is known to show only qualitative universality at low $T$ and  $c_1$ can vary for the glasses, albeit of same material, obtained by different cooling processes). 

The top of figure 3 displays a comparison of theoretically predicted $c_v^{total}$ (using eqs.(\ref{jbj7},\ref{cv1}, \ref{cvph}) in eq.(\ref{scv1})  for an amorphous basic block of Suprasil with experimental data for a macroscopic block taken from figure 6 of \cite{phil2}.  A good agreement is achieved for $T_D=190^o {\bf K}$ (used in eq.(\ref{cvph}) which is much less than that for  the macroscopic sizes of amorphous silica ($T_D \approx 495^o {\bf K}$ \cite{stephen}). 
This indicates that  (i) the super-linearity of specific heat in glasses exist even at nano scales, (ii) Debye specific heat formulation is applicable at much lower temperatures at nanoscales. 

Keeping in view the different nature of the phonon dynamics at nanoscales, an agreement of our theoretcal predictions with experimental data at macroscale is indeed a bit surprising.  We note that the prediction of a different $T_D$ for nanoscales  is consistent with previous experiments too (see e.g.\cite{nano1}). The latter also indicate that (i) the thermodynamics at nanoscales is different from macroscales, (ii) contrary to bulk materials,  $T_D$ for  nano materials can be affected by the size,  composition and dimensionality (with $T_D$ decreasing with decreasing size). Intuitively  this tendency seems to originate from a lack of sufficient strain field needed to support phonon dynamics and also their strong scattering and thereby localization at small system sizes and with increasing temperatures. However  a proper theoretical understanding of the dynamics of very high frequency phonons is still far from complete and requires further research. 

In connection with  the specific heat below $1^o \; {\bf K}$,  there have been 
experimental reports indicating an absence of linear/ super-linear behavior in some types of bulk glass materials  which has been attributed to absence of two level tunnelling states \cite{pcrrr, liu, zink, que,agl, ang}. Following  discussion  in previous section, we find that the linear temperature dependence regime of specific heat for the basic blocks depends on the competition between thermal perturbation and the mean level spacing  and is sensitive to the edge as well as bulk parameters $\lambda, b$; for the cases in which $\lambda^2 b k_b  \ll 1$ it may therefore  move to ultra low temperatures.

\vspace{0.1in}

{\bf Comparison above $T \sim 1^o \; {\bf K}$}: The bottom of figure 3 displays the theoretical $c_v^{total}/T^3$ behaviour (obtained from eqs.(\ref{jbj7},\ref{cv1}, \ref{cvph}, \ref{scv1})) along with experimental result for Suprasil taken from figure 1 of \cite{phil2}; the theoretical results, although derived  for a basic block, are almost consistent with experiments on macroscopic sizes of the glass. The bump in the heat capacity  (also referred as boson peak)  near  $T \sim 6^o \; {\bf K}$ in experimental data is however replaced by a plateau region in the theoretical result. We believe that a small deviation near the bump is a signature of the length scales of the system: contrary to semicircle DOS at nanoscales, the ensemble averaged bulk DOS at macroscopic length scales turns out to be Gaussian \cite{bb4} which is expected to lead to a boson peak in  the specific heat.

The top of figure 4 shows a comparison of  the theoretical  $c_v^{total}$ with experimental data for $a-SiO_2$ taken from figure 10 of \cite{zp}. The reasons for including  this figure here are twofold: (i) the physical properties of a glass are known to depend on the cooling process and  two different silica melts can show quantitatively different behavior (although their low $T$ properties are expected to show qualitative universality), (ii) contrary to the $c_v/T^3$ plot in figure 3, the $c_v$ plot  in figure 4 does not show a deviation from our theoretical prediction. The $c_v/T^3$ behaviour therefore seems to be more sensitive to small deviations between theory and experiments or it could be an attribute of the cooling process too. 

The bottom of figure 4 displays the comparison for another glass i.e $a-Se$;
once again the agreement with experimental data reinforces the validity of our theoretical prediction. It is important to emphasise the following difference once again:  while our theory is developed for nanosize material ($\sim 3 \; nm$), the experimental data used for comparison is obtained for macrosize sample of  e.g. $\sim 1 \; cm$. Keeping in view of a wealth of experimental information for many materials, crystalline as well as non-crystalline, indicating the deviation of physical properties at nanoscale different from macroscale (\cite{nano1} and references there in),  there is no reason to preconceive the analogy for their low temperature properties. The theoretical-experimental agreement shown in figures 3 and 4 therefore strongly hints that the origin of glass anomalies  observed abundantly in macroscopic samples is microscopic.

\section{Discussion and conclusion}

In the end, we summarize  with our main ideas and results.

The lack of detailed information due to complicated molecular interactions within a basic block  result  in  appearance  of its Hamiltonian  as a random matrix, in the product basis of single SRO molecule states; (as the property concerned here is a function of eigenvalues, it is unaffected by the choice of basis). Here the size of the block is chosen to be  the typical MRO  scale ($\sim 20-30 \AA$)  of amorphous materials where VW forces are known to dominate. At this length scale a typical amorphous material contains $\sim 10$ SRO molecules, almost all interacting with each other with same strength (by VWD forces) and homogeneously distributed within a volume of typically  $10^4 \; \AA^3$; intuitively this is expected  to lead to a delocalization of the many-body states of a block Hamiltonian. The finite mean and  variance of its matrix elements derived in section II suggest, based on the standard maximum entropy hypotheses,  the block Hamiltonian to be well-represented by a multiparametric Gaussian ensemble with its ensemble averaged spectral DOS approaching  that of a GOE (a stationary random matrix ensemble of real-symmetric matrices).  As the DOS in the spectrum edge of a GOE increases super-exponentially, this is consistent with experimental results for many body states (e.g. see \cite{redge}) and further  lends credence to the above suggestion. The higher order spectral correlations (i.e local density fluctuations) however remain close to unperturbed Hamiltonian (a summation of free SRO molecules).

We also derive the bulk DOS  for the cases in which  ${\mathcal H}$ is best described by a non-Gaussian ensemble (i.e without assuming a specific distribution); the result again agrees with the bulk GOE behavior (not surprising as semicircle DOS is known to prevail for generic distributions with finite mean and variance \cite{me}). While this does not automatically imply a GOE behvaiour in the edge too, the latter however contains very few levels (just $1$ or $2$ of them) below $e <0$, contributes to $C_v$ only at ultra low $T$ and its explicit formulation  is not needed for our analysis. Note while the full partition function of the basic block is given by eq.(\ref{jbj7}), the derivation of $C_v$ in section IV.B is based on  $\langle Z\rangle$ approximated by eq.(\ref{jbj6}). The latter corresponds to the contribution of states only from the edge-bulk boundary region. A square-root energy dependence of the DOS in this regions is necessary for a gapless spectrum and is also consistent with the edge DOS for a GOE. (Note a $\sqrt{e}$-dependence for $e>0$ in the latter follows from the Airy functions. This  is again not surprising as Airy functions are often known to appear at  boundary level problems e.g. smooth caustics in various physics domains etc). As the results derived by  the Airy function-modelling of the edge spectrum show good agreement with experiments (see figures 3, 4 for two cases and also \cite{qc1,bb3}), this lends further credence to our theoretical approach.

Based on the DOS formulation for the basic block, we find that its $C_v$ for  $T \sim 1^o K$  depends on the parameters of three competing energy scales, namely, thermal perturbation $k T$, the  bulk spectrum parameter $b$ and the edge spectral parameter $\lambda$. Based on whichever of these parameters dominates the  partition function in eq.(\ref{z1})), the  specific heat of the basic block changes from a linear to super-linear  temperature dependence. Previously observed in case of macroscopic samples, such a behavior in nano-limits was neither experimentally reported nor theoretically predicted and is a central result of our analysis.
To gain a physical insight in the origin of this behavior,  it is useful to
 note that it arises from a $\sqrt{e}$-dependence of the ensemble averaged DOS near energy $e >0$ along with a non-zero value at $e=0$ (as can be seen from eq.(\ref{jbj6}) along with  definition as $\gamma(3/2,x)=\int_0^x  \sqrt{t} \; {\rm e}^{-\beta t} \; {\rm d}t$). This is a reminder of  a similar $T$-dependence in the electronic contribution to specific heat, originating again from  a $\sqrt{e}$-dependence of the average DOS. Notwithstanding the similarity, the technical origin of $\sqrt{e}$ term is  very different in the two cases. While in the electronic case, it follows from  non-random considerations (simple counting of states),  for the block case the randomization of its Hamiltonian leads to a semi-circle density in the bulk with a super-exponentially decaying tail extended to distances of the $o(\lambda)$. Nonetheless this point deserves a deeper consideration especially in view of recent suggestions that the vibrational DOS of the amorphous system is just a modification of that of a crystal, with the BP as the broadened version of Van Hove singularity \cite{chum}.

An important aspect of  our analysis worth re-emphasizing is the size of the amorphous system used for the specific heat analysis. While previous theories   are  based on macrosize samples, the analysis  here is confined to nanosizes. This is relevant  in view of the ample evidence often indicating a variation of thermal properties from nano to macro-scales (e.g see \cite{nano1} and references therein).   Notwithstanding the different scales,  it is worth comparing  the similarities, differences and advantages of our approach with the theories for macroscales. The latter are broadly based on following basic ideas: (i)  structural disorder e.g. quasi-localized vibrations or Euclidean random matrices \cite{spm,paras, grig}, (ii) microscopic disorder in force constants (heterogeneous elasticity theory) \cite{schi1,schi2}, (iii) mesoscopic disorder in shear modulus, (iv) models based on structural correlations over distances  $10 \to 20 \AA$ (e.g \cite{du, vdos, ell3, mg}), (v) anharmonicities (e.g.\cite{gure}).

Although  the basis of our approach is instantaneous orientational disorder at the scale of medium range order,  some of our ideas seemingly overlap with previous studies.
Similar to our approach, the relevance of MRO scales to explain glass anomalies was also suggested in studies \cite{du, vdos, ell3, mg}   but these are mostly based on  existence of correlations supported by experimental/ numerical analysis and usually lack mathematical details. As discussed in \cite{bb3}, the MRO scale in our approach has a physical basis, it corresponds to the distances where phonon mediated coupling of the stress fields of two SRO molecules are balanced by the dispersion interaction. 
The heterogeneous elasticity theory \cite{schi1, schi2} predicts, similar to our case,  a standard random matrix GOE) type spectral behaviour of the vibrational states \cite{schi1} and thereby leading to the Boson peak and other glass anomalies.  But an important difference between our approach and \cite{schi1}  is as follows: the randomization of vibrational states in the basic block is not caused by static microscopic disorder but due to dynamic, instantaneous orientational disorder of the induced dipoles at the MRO scales.  Furthermore, as discussed in \cite{bb4} based on its description  as a collection of basic blocks, the Hamiltonian of a macrosize sample, in general, is predicted to be a sparse random matrix (with sparsity system dependent) and need not be a standard random matrix taken from a GOE as assumed in \cite{schi1,schi2}.

Based on description of a glass solid as elastic network, microscopic  disorder appears as a tool in  some effective medium theories too, with coordination number $z$ of the network  and compressive strain as the key parameters and neglecting large scale fluctuations of  $z$ (e.g \cite{emt}).  A generalization of these theories including weak interactions among contacts was considered in \cite{degi};  Although the latter excludes spatial elastic fluctuations, some of its results  are similar to \cite{schi1, schi2} which again suggested the irrelevance of structural disorder.  As \cite{degi} also suggests the roles of VW interactions as well  hierarchy of interactions to understand low temperature properties, this is in spirit similar to our approach. (As mentioned in section II, many types of interactions i.e electrostatic and induction (besides dispersion) may  influence the $b$-parameter in  the case of polar molecules). The results in \cite{degi} however are obtained by numerical simulations of macroscopic samples and their quantitative applicability to nano-size sample is not obvious. Further, contrary to our approach, the VW interactions in \cite{degi} are assumed to change only the energy-scaling.

Contrary to many previous theories, our theory is neither based on assumed existence of any hypothetical defects nor has any adjustable parameters. The basis of our approach are the basic structural units referred as SRO molecule, interacting by VW interactions within MRO scales and their existence is well-documented in the domain of glass chemistry (e.g see \cite{b2008}). 
The only parameter $b$ that appears in our formulation, depends  on  the molecular properties e.g. polarisability, ionisation energy, molecule volume and strengths of VW interactions which are experimentally  well-measured and a knowledge of their order of magnitudes is sufficient for our purpose; (note $\lambda$ is a constant).  As displayed in table I, $b \sim 10^{18} \; J^{-1}$  for a wide range of glasses which in turn predicts an analogous behavior of the $C_v$ at  low $T$  for them and is  consistent  with experimental observations for macro-size samples too (notwithstanding the density of states derived here  valid only for  nano-scales). 

Although the derivation of $b$ in section II.E assumes the amorphous molecules to be non-polar and thus applicable to insulators only, it can directly be extended to polar molecules by including  intermolecular  interactions of the induction type.  Furthermore our approach also suggests a possible explanation of the time-dependence of $C_v$ noted in experiments: it could arise from instantaneous aspect of the dispersion interactions.

An important question discussed extensively in the context of boson peak is its location i.e the boson peak frequency. Although the analysis presented here is confined to nanoscales, it strongly suggest the boson peak frequency $\omega_{bp}$ to be of the same order as  $e_0$, this being the edge-bulk meeting point where the DOS behavior changes from $e^{3/2}$ to a constant in $e$ (see also \cite{bb3, bb4}). (As displayed in Figure 3, the theoretical predicted $c_v^{total}$ for a nanosize $SiO_2$ sample deviates from the experimental results for a macrosize sample, only in a  small range around the bump, also referred as the boson peak of specific heat. Clearly the increased density of states in the bulk at the macroscales seems to affect $c_v^{total}$ only near the boson peak.

A lack of orientational randomization at medium range order in a crystalline material rules out  applicability of the theory discussed here  to crystals and thereby  lack of the universality in their low temperature specific heat.  But as the molecules in a crystal are also subjected to VW interactions, it is natural to query as to why the complexities of VW interactions do not lead to randomization in that case? The answer lies in various symmetries of the crystal, resulting in high number of degeneracies among the vibrational energy levels and their clustering, thus ruling out a semicircle density of states (note the latter follows due to repulsion of energy levels) and indicating, instead, a Gaussian density even in a nano-size sample.  The observed Boson peaks in crystals may then  be explained along the same route.

A complete theory of glass anomalies  is expected to explain all universalities  of low temperature physical properties. The success of our approach to explain some of them, namely specific heat, ultrasonic attenuation \cite{qc1} and Messiner-Berret ratio \cite{bb3} etc. encourages one to seek its applicability for other anomalies too; we hope to pursue some of them in near future.

\acknowledgements

I am indebted to Professor Anthony Leggett for introducing me to this rich subject and continuous intellectual support in form of many helpful critical comments  and insights  over a duration of fourteen years  in which this idea was pursued. I am also very grateful to Professor Michael Berry for advise in fundamental issues and  important technical help  in dealing with Airy function integrals which forms the basis of my calculation.   
The financial support provided by SERB, DST, India is also gratefully acknowledged.

\begin{table}
\caption{\label{tab:table I} {\bf Specific heat calculation: relevant data for 18 glasses}: here the  columns $3, 4$ give the mass density $\rho_m$ of the glass and the molar mass $M$ of the relevant unit undergoing dispersion interaction (see \cite{qc1} for details), respectively. This data is used in eq.(\ref{omb}) to obtain the volume of the basic block, displayed in column $5$. The columns $6, 7, 8, 9$ display the optical parameters for the glass, namely  refractive index $n_I$, characteristic frequency $\nu_e$, Hamaker constant $A_H$ and edge-spectral parameter $b$ (see eq.(\ref{b1}) and eq.(\ref{ah},\ref{ahs}) and \cite{sup} for calculation of $n_I$ for some of the cases; note $A_H$ used here corresponds to non-retarded dispersion interaction (see eq.(11.14) of \cite{isra} or eq.(13b) of \cite{fcdc} with medium 2 as vaccum).  From eq.(\ref{cvu}), the  specific heat for very low $T$ can be expressed as $c_v=c_1 \; T + o(T^2)$; $c_1\approx {0.22 \;\lambda \; b_0 \; k_b^2 N_a \over 64 M}$ values for 18 glasses are given in column $10$, along with available ones from \cite{stephen}. }
\begin{center}
\begin{ruledtabular}
\begin{tabular}{lccccccccccr}
Index & Glass & $\rho_m$ & $M$ & $\Omega_b$ & $n_I$  & $\nu_e$ &  $A_H$ & $b$  & $c_1$ & $c_{1,stephen}$ \\
Units & &  $\times 10^3 {Kg\over m^3}$  & ${gm \over mole}$ & $10^4 \; {angs}^3$ & & $\times 10^{15} \; sec^{-1}$ & $\times 10^{-20} \; J$ &  $\times 10^{18} \; J^{-1}$ & ${ergs\over gm-K^2}$  & ${ergs\over gm-K^2}$ \\
\hline
 1 &  a-SiO2          &    2.20 &  120.09 &    5.80 &    1.45 &    3.24 &    6.31 &    0.91 &    10 & 11\\
 2 &  BK7             &    2.51 &   92.81 &    3.93 &    1.51 &    3.10 &    7.40 &    0.78 &     12 & \\
 3 &  As2S3           &    3.20 &   32.10 &    1.07 &    2.35 &    1.41 &   17.98 &    0.32 &   14  & 14\\
 4 &  LASF            &    5.79 &  167.95 &    3.08 &    1.80 &    3.00 &   15.15 &    0.38 &    3 &\\
 5 &  SF4             &    4.78 &  136.17 &    3.03 &    1.72 &    1.98 &    8.40 &    0.68 &    7 &\\
 6 &  SF59            &    6.26 &   92.81 &    1.58 &    1.89 &    1.70 &   13.12 &    0.44 &    7 &\\
 7 &  V52             &    4.80 &  167.21 &    3.70 &    1.51 &    3.40 &    8.37 &    0.69 &    6 & \\
 8 &  BALNA           &    4.28 &  167.21 &    4.15 &    1.47 &    3.24 &    6.87 &    0.84 &   7 & \\
 9 &  LAT             &    5.25 &  205.21 &    4.15 &    1.51 &    3.79 &    9.16 &    0.63 &    4     &\\
10 &  a-Se            &    4.30 &   78.96 &    1.95 &    1.90 &    2.56 &   13.34 &    0.43 &    8  & 7.5\\
11 &  Se75Ge25        &    4.35 &   77.38 &    1.89 &    3.64 &    0.68 &   24.88 &    0.23 &    4 &\\
12 &  Se60Ge40        &    4.25 &   76.43 &    1.91 &    4.65 &    0.48 &   27.15 &    0.21 &    4 & \\
13 &  LiCl:7H2O       &    1.20 &  131.32 &   11.63 &    1.39 &    3.19 &    4.75 &    1.21 &    13   &\\
14 &  Zn-Glass            &    4.24 &  103.41 &    2.59 &    1.53 &    3.00 &    7.71 &    0.74 &    10   &\\
15 &  PMMA            &    1.18 &  102.78 &    9.26 &    1.48 &    2.82 &    6.10 &    0.94 &    13 & 46  \\
16 &  PS              &    1.05 &   27.00 &    2.73 &    1.56 &    2.15 &    6.03 &    0.95 &    49 & 51  \\
17 &  PC              &    1.20 &   77.10 &    6.83 &    1.56 &    2.08 &    6.00 &    0.96 &    17 & 38  \\
18 &  Epoxy            &    1.20 &   77.10 &    6.83 &    1.54 &    1.84 &    4.91 &    1.17 &    21  &\\
\hline
\hline
\end{tabular}
\end{ruledtabular}
\end{center}
\end{table}

\vfill\eject



\appendix

\section{Ensemble averaging of single molecule energy levels}

The consideration of the ensemble averaging in case of the single molecule energy levels, introduced in \cite{zkpcd} (also see section 4.2.2 of \cite{gmw}), can be justified on following grounds (in the regime where the levels can not be associated with exact quantum numbers): although the particle-interactions within a single molecule are not random by themselves, the missing information due to complexity of the many body interactions leads to randomization of the matrix elements of the Hamiltonian in a physically motivated basis. As a consequence, the energy levels and the eigenstates intensities can at best be described by the probability distribution. (Note similar ideas used by Wigner in  nuclear spectroscopy led to random matrix modelling of complex nuclei, atoms and molecules \cite{gmw}). The application of modern laser spectroscopy to resolve  the enormously rich and complex spectra of such molecules has indicated that the statistics for the {\it irregular} part of the spectrum can be well-described by the random matrix ensembles \cite{zkpcd} (also see section 4.2.2 of \cite{gmw} for a review of the random matrix applications of molecules). Even for the regular part (described by exact quantum numbers), the underlying symmetries result in large number of degeneracies and lead to Poisson distribution of the energy levels \cite{zkpcd,gmw}. The appearance of randomness due to underlying complexity is also supported by the detailed numerical investigations, indicating that a typical eigenstate of the molecule in a physically motivated basis (which preserves the symmetry constraints of $h_n$) is  randomly distributed, (localized or extended in the basis, subjected to symmetry constraints). As a NIM basis-state (eq.(\ref{nim1})) is a product of single molecule states, it is also expected  to be a random function.

\vspace{0.2in}

\section{Finding $e_0$ where edge meets bulk}

With bulk and the edge level densities given by eq.(\ref{rhoe1}) and eq.(\ref{rlt}) respectively, the form of the edge level density is different from that of bulk and it is  important to know how and where they connect. Let the edge meet the bulk near $e=e_0$ with $e_0$ very small, but positive energy. The latter is needed to ensure a gapless spectrum because the bulk density is zero at $e=0$ but edge-density is not. As mentioned in section III, the  point $e_0$ where edge behavior smoothly joins the bulk behavior can be given by the requirement that 
\begin{eqnarray}
\langle \rho_{edge-l}(e_0) \rangle= \langle \rho_{bulk}(e_0) \rangle
\label{e0a}
\end{eqnarray}

Using  the large $x$ behavior of Airy functions i.e $Ai(-x) \sim {1 \over \pi^{1/2} \; x^{1/4}} \; {\rm cos}\left({2 \over 3}\; x^{3/2}-{\pi \over 4}\right)$,
  $Ai'(-x) \sim {x^{1/4} \over \pi^{1/2}} \; {\rm sin}\left({2 \over 3}\; x^{3/2}-{\pi \over 4}\right)$, 
$\int Ai(-x) \; {\rm d}x \sim {1 \over 2 \pi^{1/2} \; x^{3/4}} \; {\rm cos}\left({2 \over 3}\; x^{3/2}+{\pi \over 4}\right)$, it is easy to show that 
\begin{eqnarray}
f(x) \approx {1\over \pi} \sqrt{c_0^2+ x} + {1\over \pi x} \; {\rm cos}\left({2 \over 3}\; x^{3/2}\right).
\label{rapp}
\end{eqnarray}
The above gives $\langle \rho_{edge-l}(e) \rangle \approx {\xi b_0 \over \pi} {\sqrt{c_0^2+ \lambda b_0 e}}$ for $e >0$. 
Our nest step is to figure out what is $\xi$ and $\lambda$? For this we note that the functional form of both $\rho_{bulk}(e)$ and  $\rho_{edge-l} (e)$ should match near $e \sim e_0$.
Substitution of the latter along with eq.(\ref{rhoe1}) in eq.(\ref{e0a}) gives 
\begin{eqnarray}
\xi \sqrt{c_0^2 + \lambda b_0 e_0} =  \sqrt{2 b_0 e_0}
\label{e0b0}
\end{eqnarray}
Using $\xi \sqrt{\lambda} =1$ and $c_0^2=1/3$ in the above, we have $ e_0 =  {c_0^2 \over b_0 \lambda}  \approx {1 \over 3 b_0 \lambda} $.

To find out exactly the point $e_0$ i.e where $f(x)$ in eq.(\ref{a1l}) begins to behave as a square-root, we take help of Mathematica which gives $ \lambda b e_0 \approx 3$ and therefore $e_0 \approx  {1 \over 3 \lambda  b_0}$ (See also figure 1).

\section{Superlinear $T$-dependence}

The difference between two $\gamma$-functions appearing in eq.(\ref{jbj6}) becomes exponentially smaller with increasing $\beta$. Although this is balanced by the presence of  expoential term therein,  the different series  approximations of $\gamma$-functions mentioned below eq.(\ref{gm1}) are no longer appropriate.  Although technically complicated, a better approximation of $C_v$ can be given as follows. Taking $\log$ of  eq.(\ref{jbj6}) and using the approximation $\log(1+y) \approx y +{y^2\over 2}+....$, 
\begin{eqnarray}
\log \langle Z \rangle  &\approx & 
{\sqrt{\eta^3}\over  \pi \sqrt{27 \; \lambda_0^3}} \; 
\; \left({1\over x }\right)^{3/2}  \; \left( \Gamma\left({3\over 2},  {x}\right) - \Gamma\left({3\over 2},  {(1+\eta^{-1}) x}\right) \right) \; {\rm e}^{x}  
\label{zbb3b}
\end{eqnarray}
with $\Gamma(a,x)=1-\gamma(a,x)$ and $x=\eta \beta e_0$. Further using the relation  ${\partial   \; \Gamma\left(a, x\right) \over \partial x} =  - x^{a-1} \;  {\rm e}^{-x}$,  eq.(\ref{cv1}) then gives
\begin{eqnarray}
C_{v}(T)  &\approx  &
 {k_b \sqrt{2 \eta^3}\over \pi \sqrt{27 \; \lambda_0^3}  } {1\over x^{3/2}} \left[\left(15/4-3 x +  x^2 \right)\left( \Gamma\left({3\over 2},  {(1+\eta^{-1}) x}\right) - \Gamma\left({3\over 2},  {x}\right) \right) {\rm e}^{x} +\right . \nonumber \\ 
 && \left . - {(1+ \eta)^{3/2} \over 2 \eta^{5/2}} \left(6 \eta -\eta x + 4 x^2 \right)  x^{1/2}  {\rm e}^{-x/\eta} +{1\over 2} x^{3/2} (1-x)  \right]
\label{cvvb}
\end{eqnarray}
The power of the leading term of $C_v(T)$ can now be seen by calculating the local power law exponent $\alpha ={ {\rm d}\log C_v \over {\rm d} \log T}$. Following from eq.(\ref{cvvb}), $\alpha$ varies between $1\to 1.5$ as $T$ varies.


\newpage
 
\begin{figure}[ht!]
\centering
\includegraphics[width=1.\textwidth]{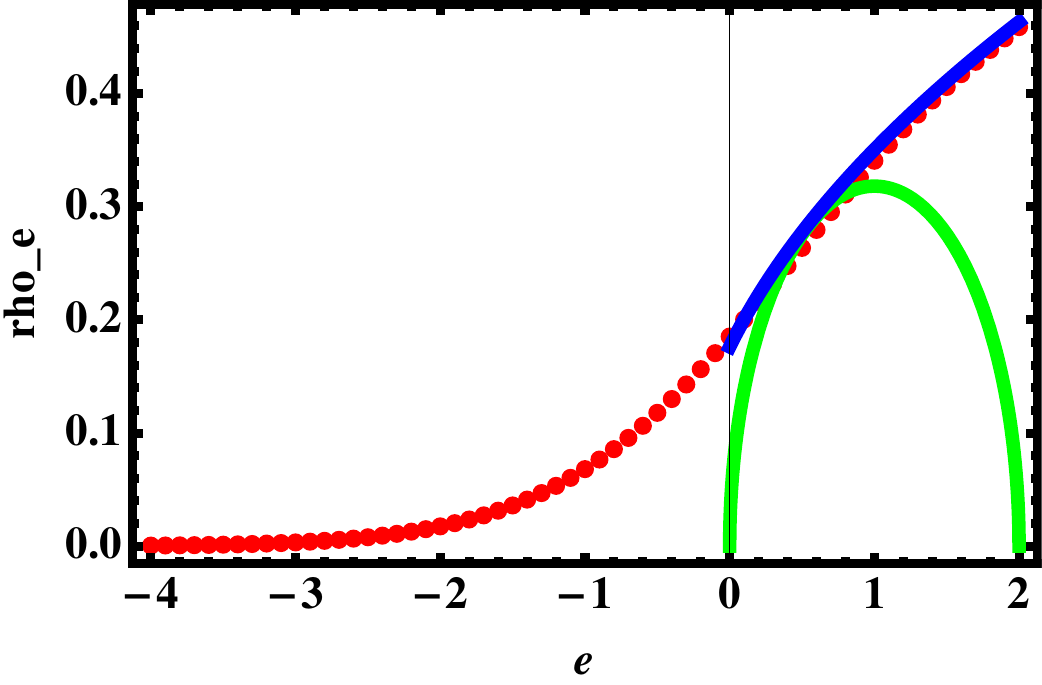}
\caption{
{\bf Finding $e_0$ where edge meets bulk?:} Bulk density of state $\rho_{bulk}(e)$ (eq.(\ref{rhoe1}), green color) and the edge density of states $\rho_{edge}(e)$ (eq.(\ref{rlt}) with $f(x)$ given by eq.(\ref{a1l}), red dots). The approximation (\ref{rhop}) of $\rho_{edge}(e)$ for $e >0$ is also depicted by blue color.}
\label{fig1}
\end{figure}

\begin{figure}[ht!]
\centering
\includegraphics[width=\textwidth]{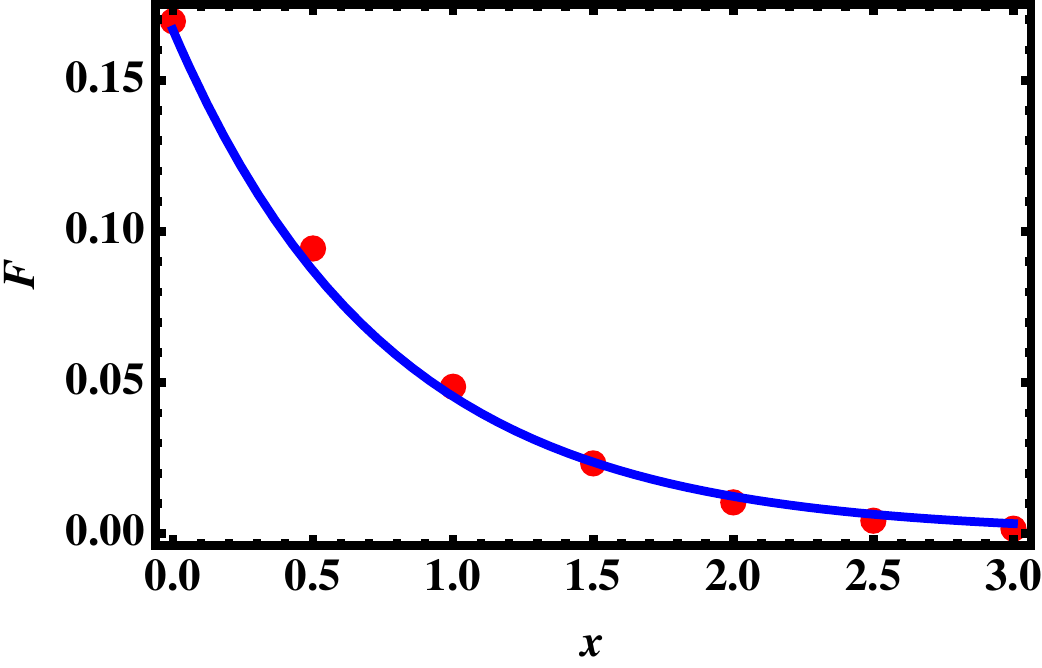}
\caption{
{\bf Counting function ${\mathcal F}(x)=\int_{-\infty}^{x} \; {\rm d} x \;  f_L(x)$ (see eq.(\ref{nlt})):  } A rapid exponential decay of ${\mathcal F}(\lambda b_0 e)$ (red dots) away from $e=0$ indicates most of the eigenvalues lie close to $e=0$. The solid curve (blue color) describes the fit $(1/6) {\rm exp}[-1.3 \; x]$.}
\label{fig2}
\end{figure}

\begin{figure}[ht!]
\centering
\includegraphics[width=\textwidth]{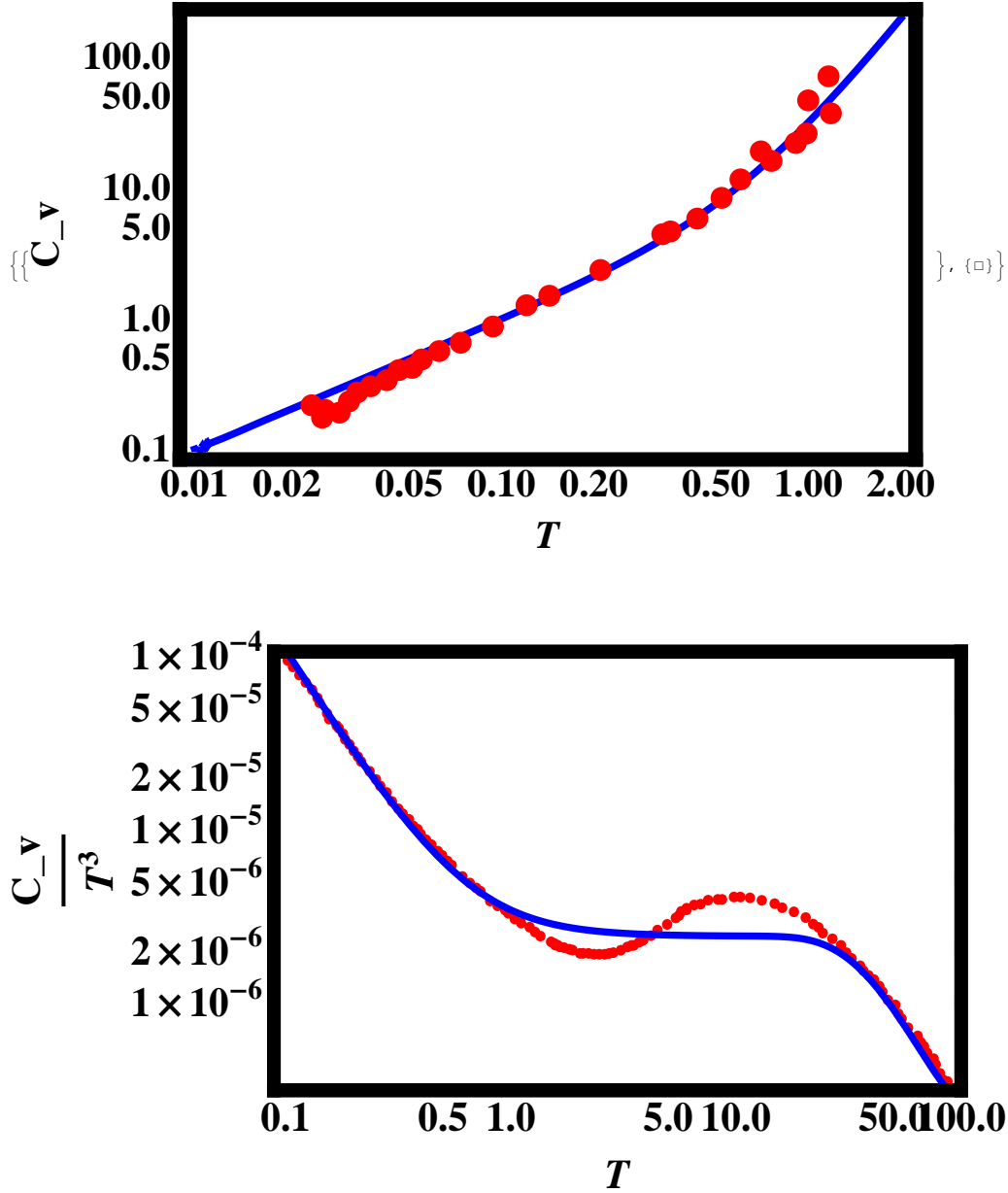}
\vspace{-1.5in}
\caption{
Comparison of total specific heat $c_v^{total}$ of a nanosize basic block with experimental data for a macrosize sample: $c_v^{total}=c_v+c_v^{ph}$, with $c_v={1\over \rho_m \Omega_b} \; \langle C_v \rangle$, obtained directly by substituting  eq.(\ref{jbj7}) in eq.(\ref{cv1}) and $c_v^{ph}$ from eq.(\ref{cvph})  is displayed by the blue curve. Experimental data for Suprasil (red curve) is taken  from figure 6 of \cite{phil2}  for {\it Top} case and from figure 1 of \cite{phil2} for {\it bottom} case. The $T_D$ value which gives a better fit in this case is much less than that of bulk;  $T_D$ used  in eq.(\ref{cvph})  is  $190^o {\bf K}$ while $T_D \approx 495^o {\bf K}$ for the macroscopic sizes of $a-SiO_2$ \cite{stephen}. This is however not unexpected (see the end of section IV).
}
\label{fig3}
\end{figure}

\begin{figure}[ht!]
\centering
\includegraphics[width=\textwidth]{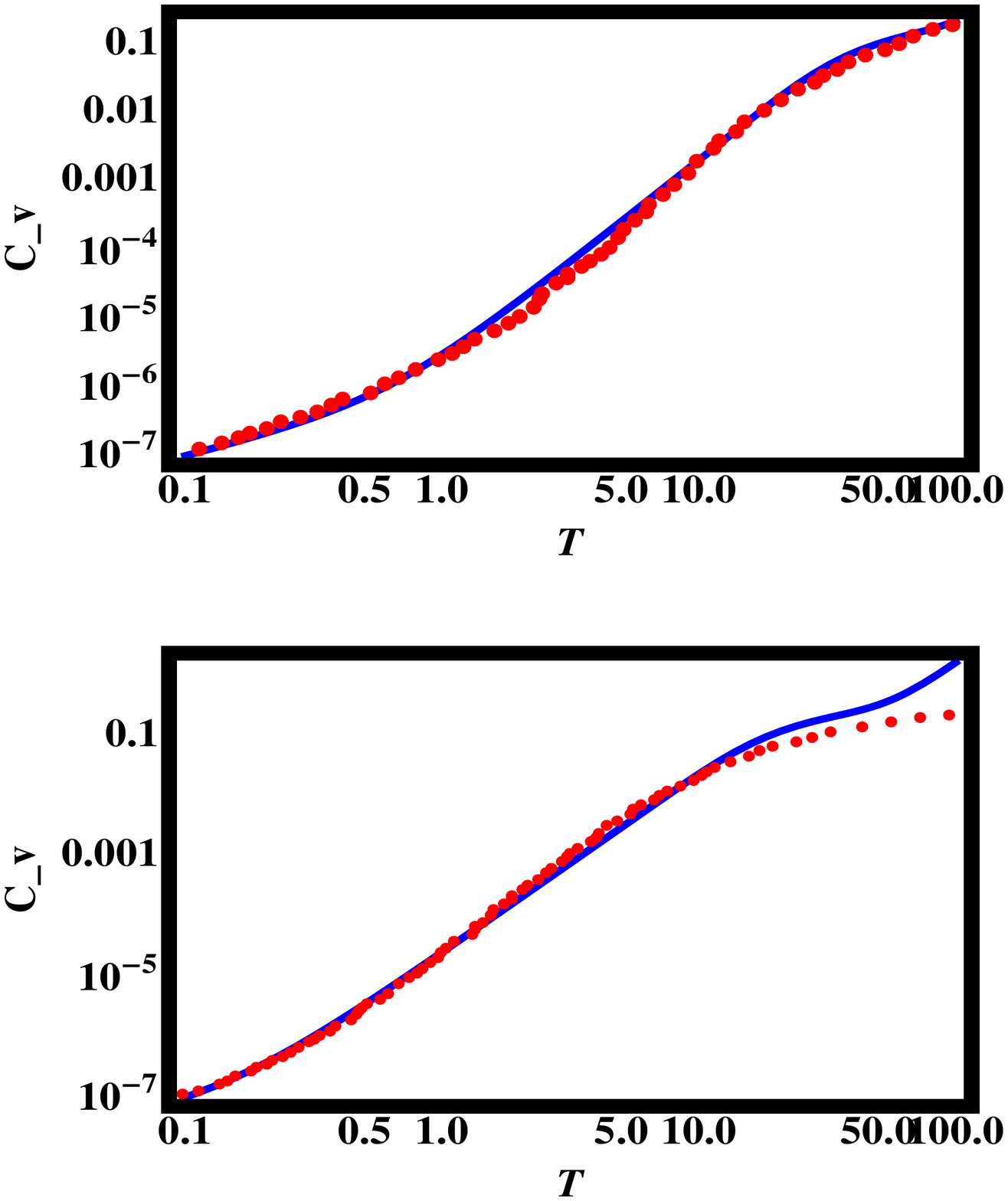}
\vspace{-1.5in}
\caption{
Comparison of total specific heat  $c_v^{total}$ for a nanosize sample with experimental data for a macrosize sample of $a-SiO_2$ (different from Suprasil, another Silica melt, in figure 3) and $a-Se$: $c_v^{total}=c_v+c_v^{ph}$, with $c_v$ obtained directly by substituting  eq.(\ref{jbj7}) in eq.(\ref{cv1})  and $c_v^{ph}$ from eq.(\ref{cvph}), is displayed by the blue curve.  {\it Top}:  $a-SiO_2$ (with $T_D=200^o {\bf K}$), {\it Bottom}: $a-Se$ (with $T_D=113^o {\bf K}$). Experimental data (red curve) for $a-SiO_2$ and $a-Se$ is taken from figure 10 and figure 12 of \cite{zp}, respectively. 
Note although dependent on the origin of material, typically $T_D \approx 495^o {\bf K}$ and $123^o {\bf K}$  for the macroscopic sizes of $a-SiO_2$ and $a-Se$\cite{stephen}. 
}
\label{fig4}
\end{figure}


\newpage


\begin{thebibliography}{10}


\bibitem{zp}
R.C. Zeller and R.O. Pohl, Phys. Rev. B, 4, 2029, (1971).

\bibitem{stephen}
R. B. Stephens, Phys. Rev. B, 8, 2896, (1973).

\bibitem{plt}
R.O.Pohl, X.Liu and E.Thompson, Rev. Mod. Phys. 74, 991, (2002). 

\bibitem{and}
P.W. Anderson, B.I. Halperin and C.M. Verma, Philos. Mag. 25, 1, (1972). 

\bibitem{phil}
W.A. Phillips, Two Level States in Glass, rep. Prog. Phys. 50, 1657, (1987);
R. Hunklinger and K. Raychandharai, in Progr. Low-Temp. Phys. (ed. D. F. Brewer, Elsevier, Amsterdam), 9, 265, r1986;
J. Jackle ,  {\it Amorphous Solids: Low-Temperature Properties}, (Springer, Berlin) 1981.

\bibitem{galp}
Y. M. Galperin, V. G. Karpov and N. Solovjevv,  Eksp. Teor. Fiz., 94 (1988) 373.

\bibitem{ram}
M.A. Ramos, Low Temp. Phys. 46, 104 (2020).

\bibitem{rpjr}
M. A. Ramos, T. Pérez-Castañeda, R. J. Jiménez-Riobóo, C. Rodríguez-Tinoco, and J. Rodríguez-Viejo, Low Temp. Phys. 41, 412 (2015);


\bibitem{lg1}
A. J. Leggett and D. Vural, J. Phys. Chem. B, 42,117, (2013).



\bibitem{lg2}
A.J. Leggett, Physica B: Cond. Matt. 169, 332 (1991).



\bibitem{yl}
C.C.Yu and A.J.Leggett, Comments Condens Matter Phys 14, 231, (1988).


\bibitem{pcrrr}
T. Perez-Castaneda, C. Rodriguez-Tinoco, J. Rodriguez-Viejo and M. A. Ramos, PNAS, 111, 11275, (2014).  

\bibitem{liu}
Xiao Liu, B. E. White, Jr., R. O. Pohl, E. Iwanizcko, K. M. Jones, A. H. Mahan, B. N. Nelson, R. S. Crandall, and S. Veprek, Phys. Rev. Lett., 78, 4418, (1997).

\bibitem{zink}
B.L. Zink, R. Pietri and F. Hellman, Phys. Rev. Lett., 96, 055902, (2006).

\bibitem{que}
D. R. Queen, X. Liu, J. Karel, T.H. Metcalf and F. Hellman, Phys. Rev. Lett., 110, 135901, (2013).

\bibitem{agl}
N.I. Agladze, A.J. Sivers, Phys. Rev. Lett., 80, 4209, (1998). 

\bibitem{ang}
C. A. Angell, C.T. Moynihan and M. Hemmati, J. Non-Crystalline solids, 274, 319, (2000).




\bibitem{spm}
V.G.Karpov, M.I.Klinger, F.N.Ignatiev, Sov. Phys. JETP 57, 439, (1983).

\bibitem{buch}
U. Bucheanau, Y. M. Galperin, V. Gurevich, D. Parashin, M. Ramos and H. Schober, Phys. Rev. B 46, 2798, (1992); 
43, 5039, (1991)


\bibitem{paras}
D. A. Parashin, Phys. Rev. B, 49, 9400, (1994). 



\bibitem{gure}
V. Gurevich, D. Parashin and H. Schrober, Phys. Rev. B, 67, 094203, (2003). 


\bibitem{schi1}
W. Schirmacher, G. Diezemann and Carl Ganter, Phys. Rev. Lett.,81, 136, (1998);
W. Schirmacher, Euro. Phys. Lett., 73, 892, (2006).

\bibitem{schi2}
 A. Maruzzo, W. Schirmacher, A. Fratalocchi and G. Ruocco, Sci. Rep., 3, 1407, (2013).


\bibitem{grig}
T. Grigera, V. Martin-Mayor, G. Parisi and P. Verrocchio, Nature, 422, 289, (2003). 

\bibitem{emt}
M. Wyart, Euro. Phys. Lett., 89, 64001, (2010). 

\bibitem{degi}
E. DeGiuli, A. Laversanne-Finot, G. During, E. Lerner and M. Wyart, Soft Matter, 10, 5628, (2014).


\bibitem{ell3}
S.R.Elliott, Europhys. Lett. 19, 201 (1992).

\bibitem{du}
E. Duval, A. Boukenter, T. Achibat,  J. Phys. Condens. Matter 2, 10227, (1990). 



\bibitem{vdos}
V. K. Malinovsky, V. N. Novikov, P.P. Parashin, A.P. Solokov and M.G. Zemlyanov, Europhys. Lett., 11, 43 (1990). 


\bibitem{mg}
G. Monaco and V. M. Giordano, PNAS.0808965106.



\bibitem{yu}
C.C.Yu, Phys. Rev. Lett., 63, 1160, (1989). 


\bibitem{vl}
D. Vural and A.J.Leggett, J. Non crystalline solids, 357, 19, 3528, (2011). 


\bibitem{vlw}
V. Lubchenko and P. G. Wolynes, Phys. Rev. Lett. 87, 195901, (2001). 

\bibitem{misha}
M. Turlakov, Phys. Rev. Lett., 93, 035501, (2004). 

\bibitem{burin}
A.L. Burin, J. Low. Temp. Phys. 100, 309 (1995);
A.L.Burin and Y.Kagan, Physics Letters A, 215 (3-4), 191, (1996).


\bibitem{Seth}
J. P. Sethna and K.S. Chow, Phase Tans. 5, 317 (1985); 
M. P. Solf and M.W.Klein, Phys. Rev. B 49, 12703 (1994).



\bibitem{zac}
M. Baggioli and A. Zaccone, Phys. Rev. Research 1, 012010(R), (2020); 
M. Baggioli, R. Milkus, and A. Zaccone, Phys. Rev. E 100, 062131, (2019).


 

\bibitem{nano1} 
J.P.Setrajcic, V.D.Sajfert, S.K.Jacimovski, Review in Theoretical Sciences, DOI: 10.1166/.2016.1067, 2016);
L. Qiu, N. Zhu, Y. Feng, E. E. Michaelides, G. Żyła,
D. Jing, X. Zhang, P. M. Norris, C. N. Markides, O. Mahian, Physics Reports 843, 1, (2020).



\bibitem{phil2}
J.C. Phillips, J. Non-crys. solids 43, 37, (1981); 34, 153, (1979). 

\bibitem{ell1}
S.R.Elliott, Nature, 354, 445, (1991).

\bibitem{sup} P. Shukla, Supplementary material.

\bibitem{sme} A. Smekal,  J. Soc. Glass Technol., 35, 411, (1951).

\bibitem{b2008}
D.Hulsenberg, A. Harnisch, A. Bismarck, {\bf Microstructuring of Glasses}, ed. Springer, (2008).

\bibitem{b1991}
H. Scholze, {\bf Glass: nature, structure and properties}, Springer-Verlag, (1991).

\bibitem{b1999}
H. Bach and D. Krause, {\bf Analysis of the Composition and Structure of Glass and Glass Ceramics}, Springer, (1999).

\bibitem{b1994}
W.Vogel, {\bf Glass Chemistry}, 2nd Ed.,Springer-Verlag, (1994).


\bibitem{mad}
A. Aguado, P.A. Madden, Phys. Rev. B 70, 245103 (2004);
M. Salanne, B. Rotenberg, S. Jahn, R. Vuilleumier, C. Simon and P. A. Madden; arXiv:1204.1427v1.

\bibitem{ajs}
A.J.Stone, {\it The theory of intermolecular forces}, Oxford scholarship online, Oxford university Press, U.K. 2015.


\bibitem{sbb}
Sorensen et al, Sci. Adv.6:eabc2320, (2020).


\bibitem{cheng}
S. Cheng, Ceramics, 4, 83. (2021).


\bibitem{tang1} 
K. T. Tang  and J. P. Toennies, J. Chem. Phys., 118, 4976, (2003); 80, 3726, (1984); 
S. Warnecke, K. T. Tang  and J. P. Toennies,  J. Chem. Phys.,  142, 131102, (2015).

\bibitem{bks}
B. van Beest, G. Kramer, and R. van Santen. Phys. Rev. Lett., 64:1955, 1990.

\bibitem{ttam}
S. Tsuneyuki, M. Tsukada, H. Aoki, and Matsui. Phys. Rev. Lett., 61:869, 1988;




\bibitem{exp1}
The NIM basis can be truncated for the following reason. The typical energy of excitation for electronic, vibrational and rotational level in molecules is of the order of $10^{-19} \; J, 10^{-20} \; J, 10^{-23} \; J$ respectively. The thermal perturbation at very low temperatures ($T < 30 K$) is not strong enough to result in  the electronic state excitations and  the molecule remain in its electronic ground state. It could however cause transitions to its roto-vibrational excited states. 



\bibitem{gmw}
T. Guhr, G. A. Muller-Groeling and H. A. Weidenmuller, Phys. Rep. 299, 189 (1998).

\bibitem{me}
M.L.Mehta,{\it  Random Matrices}, Academic Press, (1991). 



\bibitem{zkpcd}
Th. Zimmerman, H. Koppel, L. S. Cederbaum, G. Persch and W. Demtrbder, Phys. Rev. Lett. 61 3, (1988). 



\bibitem{note1}
Here $(C^{(3n)}_{\mathcal KL})^2$ is first replaced by ${\mathcal  C}_{3n}^2$ which then permits the replacement of  the sum $\sum_{\mathcal L=1}^{N} $ by the equivalent sums  $\sum_{n=2}^{g_0}  \; \sum_{q1,..,qn =1 \atop q1 \not= q2 ..\not=qn}^{g_0} $; the latter sum here corresponds to, for a given $n$, to all $n$-molecules which undergo transition from their respective single-particle states in ${\mathcal K}$ to those in ${\mathcal L}$.


\bibitem{qc1} P. Shukla, arXiv:2009.00556.


\bibitem{bb1} P. Shukla, arXiv:2008.12960.
\bibitem{bb3} P. Shukla, arXiv:2101.00492
\bibitem{bb4} P. Shukla, arXiv:2102.11216



\bibitem{kr}I.G.Kaplan and O.B. Rodimova, Sov. Phys. Usp. 21(11), 1978. 




\bibitem{cn}
For qualitative analysis, ${\mathcal C}_6$ can  be approximated by London's formula
${\mathcal C}_6 \approx {3\over 2} \alpha^2 I$  with $\alpha$ as the molecular polarizability averaged over all orientations and $I$ as the first ionization potential \cite{more, ajs, isra, kr}). (As mentioned in chapter 6 of \cite{ajs}, $\alpha$ is roughly proportional to molecular volume and ionization potentials do not differ much among different molecules). Similarly ${\mathcal C}_9 \approx {3/4} \alpha \; {\mathcal C}_6$ (applying eq.(10.2.5) of \cite{ajs} for identical molecules). Taking typical values of the molecular polarizability $\alpha \sim 10^{-30} \; m^3$ and $I \sim 10^{-18} \; J$, gives ${\mathcal C}_6 \sim 10^{-78} \; J-m^6$ and ${\mathcal C}_9 \sim 10^{-108} \; J-m^9$. 


\bibitem{more} 
Although more exact expressions for $C_6$ and $C_9$ are also available e.g  see chapter 4 of \cite{ajs}) or eq.(2.34) to eq.(2.40) of \cite{kr} but London's formula gives fairly accurate values for interactions in a vaccum. These values are usually lower than more rigorously determined ones. 

\bibitem{isra}
J. Israelachvili, Chapter 11, {\it Intermolecular and Surface Forces}, 3rd ed. Academic Press, (2011).


\bibitem{fr2000}
R.H. French, J. Am. Ceram. Soc., 83, 2117, (2000).




\bibitem{hjs}
H-J Stockmann, {\bf Quantum Chaos: an introduction}, Cambridge univ. Press (1999) (see page 79). 

\bibitem{issr}
 L. Isserlis, Biometrika, 12, 134, (1918).
 

\bibitem{bray}
G. J. Rodgers and A. J. Bray, Phys. Rev. B, 37, 3557, (1998); A. Khorunzhy and G. J. Rodgers, J. Math. Phys. 38, 3300 (1997).


\bibitem{thou}
R.C Jones, J M Kosterlitz and D J Thouless,  J. Phys. A: Math. Gen., 11, 3,1978.

\bibitem{airy}
P. Forrester, Nucl. Phys. B, 402, 709, (1993).



 

\bibitem{fcdc}
R.H.French, R.M.Cannon, L.K.DeNoyer and Y.-M. Chiang, Solid State Ionics, 75, 13, (1995);



\bibitem{dive}
In case of the standard formulation for the partition function, (with ground state assumed to be at $e=0$), the lower and upper limits are $e=0$ and $\infty$, respectively. To avoid divergence of the integral at negative energies, extra care for the lower limit must be taken for the cases where the ground state occurs at $e <0$. Following from eq.(\ref{a1l}), although the theoretical form of the density of states in our case extends to $e \to -\infty$, however as discussed in section III, very few i.e $1-2$ levels exist in  the region $e \sim 0$ even in large $N$ limit (with $N$ as size of the spectrum). Using a continuous form of the $\rho(e)$ is therefore not justified in this region and can introduce errors in the calculation. 




\bibitem{redge}
M. G. Vavilov, P. W. Brouwer, V. Ambegaokar, and C. W. J. Beenakker
Phys. Rev. Lett. 86, 874, (2001).


\bibitem{lern}
E. Lerner, G. Düring, and E. Bouchbinder, Phys. Rev. Lett., 117, 035501, (2016).

\bibitem{mizu}
H. Mizuno, H. Shiba and A. Ikeda, PNAS 114, E9767, (2017). 

\bibitem{chum}
A. I. Chumakov et al., Phys. Rev. Lett. 106, 225501 (2011).



\bibitem{wang}
Y. Wang, L. Hong, Y. Wang, W. Schirmacher and J. Zhang, Phys. Rev. B 98, 174207 (2018). 
 
\bibitem{nie}
 Y. Nie, H. Tong, J. Liu, M. Zu, N. Xu, Front. Phys. 12, 126301 (2017)


\bibitem{psall} 
P. Shukla, J. Phys.: Condens. Matter 17, 1653 (2005); Phys. Rev. E, 62, 2098, (2000).

\bibitem{psco}
P. Shukla, Phys Rev. E 71, 026226 (2005); J. Phys. A: Math. Theor. 41, 304023 (2008).



\end{thebibliography}
\end{document}